\documentstyle[aps,prb,epsf,multicol]{revtex}
\begin{document}
\draft
\title{Vortex dynamics for two-dimensional $XY$ models}
\author {Beom Jun Kim, Petter Minnhagen, and Peter Olsson}
\address {Department of Theoretical Physics, Ume{\aa} University,
901 87 Ume{\aa}, Sweden}
\preprint{\today}
\maketitle
\begin{abstract}

Two-dimensional $XY$ models with resistively shunted junction (RSJ) dynamics
and time dependent Ginzburg-Landau (TDGL) dynamics are simulated and it is
verified that the vortex response is well described by the
Minnhagen phenomenology for both types of dynamics. Evidence is
presented supporting that the dynamical critical exponent $z$ in the low-temperature phase is
given by the scaling prediction (expressed in terms of the Coulomb gas temperature
$T^{\rm CG}$ and the vortex renormalization given by the dielectric constant
$\tilde\epsilon$) 
$z=1/\tilde{\epsilon}T^{\rm CG}-2\geq 2$ both for RSJ and TDGL and
that the nonlinear $IV$ exponent $a$ is given by $a=z+1$ in the low-temperature phase. 
The results are discussed and compared with the results of
other recent papers and the importance of the boundary
conditions is emphasized. 

\end{abstract}


\begin{multicols}{2}

\section{Introduction} \label{sec_intro}

Superconducting films and two-dimensional (2D) Josephson junction
arrays as well as $^4$He films undergo Kosterlitz-Thouless (KT) type
transitions from the superconducting/superfluid to the normal
state.~\cite{kosterlitz,minnhagenrev}
The KT transition is driven by thermally created vortex-antivortex
pairs which start to unbind at the transition.~\cite{minnhagenrev}
This means that some dominant characteristic features of the physics close to
the transition are associated with vortex pair fluctuations.
The great current interest in 2D vortex fluctuations stems
from the fact that they are  
also present in 
high-$T_c$ superconductors, not only in the case of thin films,
but also in 3D samples 
just above the transition.~\cite{varenna}
It is therefore of interest to understand the properties associated
with
these thermally created vortices. 
Whereas there is a fairly good consensus on the  static properties
associated with vortex pair fluctuations,~\cite{varenna}
the dynamical aspects are less clear and some features are still
controversial.

The knowledge of the dynamical properties of vortex fluctuations
mainly comes
from experiments on superconducting
films and $^4$He films,~\cite{minnhagenrev,varenna} and from various model simulations.~\cite{varenna}
The theoretical attempts are so far on a rather phenomenological level~\cite{minnhagenrev,ambegaokar,bormann} 
with few exceptions.~\cite{capezalli}
The more explicit knowledge derives from  several kinds of simulations: $XY$ models
with time dependent Ginzburg-Landau (TDGL) dynamics,~\cite{jonsson}
$XY$ models with resistively shunted 
Josephson junction (RSJ) dynamics,~\cite{tiesinga,simkin} the Coulomb gas model with
Langevin dynamics,~\cite{holmlund}
and the lattice Coulomb gas model with Monte Carlo
dynamics.~\cite{weber} There exist two
phenomenological descriptions: the Ambegaokar-Halperin-Nelson-Siggia
(AHNS) description~\cite{ambegaokar}
and the Minnhagen phenomenology (MP).~\cite{minnhagenrev} 
There are, likewise, two distinct
proposals for the nonlinear $IV$ exponent $a$, i.e.,
$a_{\rm AHNS}$ (Ref.~\onlinecite{ambegaokar})
and $a_{\rm scale}$ (Ref.~\onlinecite{minnhagen1}) with a corresponding proposal
for a critical dynamical exponent $z=a_{\rm scale}-1$ (Ref.~\onlinecite{minnhagen1})
in the low-temperature phase. It has also been argued that the
nonlinear $IV$ exponent with the value $a_{\rm scale}$ applies to an
intermediate current range whereas $a_{\rm AHNS}$ should be recovered
in the true small-current limit.~\cite{bormann} This argument rests on
the assumption that for any finite current there are free vortices
present and furthermore that these free vortices can be described by a
conventional dynamics with $z=2$.~\cite{bormann}

In this paper we present extensive
simulations of 2D $XY$ models with RSJ as well as TDGL dynamics  using an
unconventional boundary condition. This enables us to obtain more
information on the vortex dynamics for these models.

The situation is roughly the following: The MP form of the
dynamical response gives a good description of the 2D $XY$ models with
TDGL dynamics,~\cite{jonsson} the Coulomb gas model with Langevin
dynamics,~\cite{holmlund}
and experiments on 2D superconductors.~\cite{jonsson,theron,rogers}
In the present paper we show that it also gives a good description of
2D $XY$ models with RSJ dynamics. The dynamical exponent $z$ for the
lattice Coulomb gas
with Monte Carlo dynamics has from simulations been inferred to have
the scaling value
$z=a_{\rm scale}-1$.~\cite{weber} In the present paper we verify this
result for the $XY$ models with both RSJ and TDGL dynamics. This is
seemingly
in contradiction to the results in Ref.~\onlinecite{tiesinga} that
the 2D $XY$ models with RSJ and TDGL dynamics behave differently and
appear
to have different $z$ values. The nonlinear $IV$
exponent
$a$ has been found to have the scaling value $a_{\rm scale}$ for the
Coulomb gas with Langevin dynamics~\cite{holmlund} and the lattice Coulomb gas 
with Monte Carlo dynamics.~\cite{weber} However,
contradictory    
results have been found for the $XY$ model with RSJ dynamics,
e.g., $a=a_{\rm AHNS}$ in Ref.~\onlinecite{simkin} and $a=a_{\rm scale}$ in
Ref.~\onlinecite{minnhagen1}.
In the present paper we find support for $a=a_{\rm scale}$ for the 2D $XY$ model
with RSJ dynamics.

The picture emerging from our perspective is a generic vortex response well described
 by the MP form
 of the frequency response, the scaling exponent $a_{\rm scale}$ and the
 corresponding
 dynamical exponent $z=a_{\rm scale}-1$. According to our view this
 generic vortex response
 describes both Coulomb gas models and 2D $XY$ models and is 
 insensitive to the detailed type of the dynamics be it Coulomb gas Langevin-, Monte Carlo-,
 TDGL-, or RSJ-type. 

The content of the present paper is the following: In Sec.~\ref{sec_XY} we
describe the $XY$-type models and the relevant correlation and response functions,
as well as the relation to the vortex and Coulomb gas degrees of
freedom. We also discuss the validity of linear response and
the relation between the complex impedance
and the dielectric function of the Coulomb gas.
In Sec.~\ref{sec_dyn} the dynamical equations are described and the
boundary condition is introduced and discussed. Sections~\ref{sec_result} and \ref{sec_nonlinear} 
contain
our simulation results; Sec.~\ref{sec_result} the equilibrium ones and Sec.~\ref{sec_nonlinear}
the result when the system is driven by an external current. Finally
in Sec.~\ref{sec_final} we summarize our results and make some final remarks.

\section{$XY$ Model} \label{sec_XY}

On a phenomenological level, a 2D superconductor/superfluid
can be described by an order parameter $\psi({\bf r})=|\psi({\bf
  r})|e^{i\theta({\bf r})}$, where $|\psi({\bf r})|^2$ is proportional to the
superfluid 
density and ${\bf \nabla}\theta ({\bf r})$ is proportional to the superfluid
velocity.~\cite{minnhagenrev} The energy associated with the order
parameter is the kinetic energy of the current and consequently the
energy is proportional to 
$\int d^2r[{\bf \nabla}\theta ({\bf r})]^2/2$.~\cite{minnhagenrev} A positive
(negative) vortex centered at a certain point is associated with the topological excitation characterized by that
the line integral $\int {\bf \nabla}\theta({\bf r})\cdot d{\bf l}$ of
an arbitrary
small closed loop around the point is equal to $2\pi$ ($-2\pi$). There is
a precise mapping between the vortices of a 2D superconductor and 2D
Coulomb gas charges.~\cite{minnhagenrev} Since our interest in the
present paper is the dynamical effects of the thermal vortex
fluctuations, we will describe our results in the language of 2D Coulomb gas charges. 

The $XY$-type models in a broad sense are models representing
the continuum order parameter $\psi({\bf r})=|\psi({\bf
  r})|e^{i\theta({\bf r})}$ put on a lattice.    
Let us for convenience
choose a square lattice. The discretized
version is then $\psi_j=|\psi_j|e^{i\theta_j}$, where the index
$j$ denotes the lattice points. Let us simplify further by neglecting the
variations of the magnitude of the order parameter and take
$|\psi_j|=|\psi|$ to be a constant. The discretized version of the
energy then takes the form
\begin{equation}
  H_{XY}=J\sum_{\langle ij\rangle}U(\phi_{ij}=\theta_i-\theta_j) ,
  \label{xy}
\end{equation}
where $J\propto|\psi|^2$ is termed the $XY$ coupling constant
and the sum is over
nearest-neighbor pairs. The lattice constant is taken to be unity so
that $\phi_{ij}=\theta_i -\theta_j$ corresponds to $\nabla
\theta$ (in the direction from $j$ to $i$). The function $U(\phi)$
has to be equal to $\phi^2/2$ for small $\phi$ in order to
yield the correct continuum limit and in addition $U(\phi )$ has to be
a periodic function of $2\pi$ since the phase angle $\theta_i$ for
each lattice point is only defined upto a multiple of $2\pi$.
A possible choice for $U(\phi )$ is then
\[ U(\phi )=1-\cos\phi  \]
and with this choice the model is the usual 2D
$XY$ model or the planar rotor model. This particular interaction would,
e.g., arise if each lattice point was a small superconducting island
which was Josephson coupled to its nearest neighbors, and the
system is called
a Josephson junction array (JJA). We will use this
choice of the interaction in the present paper. However, from
the point of view of vortex fluctuations any $U(\phi )$ fulfilling the
necessary requirements stipulated above is a valid choice.
A possible generalization is 
\begin{equation}
  U(\phi)=\frac{2}{p^2}\left[ 1 -\cos^{2p^2}\left(\frac{\phi}{2}\right)\right] ,
\label{pmodel}
\end{equation}
where $p=1$ corresponds to the usual
$XY$ model. The practical point with such a generalization is that the
vortex density increases with increasing $p$.~\cite{domany}
Consequently the vortex response is sometimes easier to extract from
simulations for a $p$ value larger than 1.~\cite{jonsson}

The Boltzmann factor for a particular configuration is given by
$e^{-H_{XY}/T}$ where $T$ is the temperature in units of $k_B=1$. 
From this all thermodynamic properties can be obtained.

The mapping between the $XY$ model and the Coulomb gas representation is
as follows:~\cite{olsson} The effective temperature variable for the Coulomb gas
charges is given by  $T^{\rm CG}=T/[2\pi J\langle U''\rangle]$, where $T$
is the temperature for the $XY$ model, $\langle \cdots  \rangle$ denotes a
thermal average, and $U''={\partial}^2 U / \partial \phi^2$. The supercurrent through a link 
is given by
$JU'=J\partial U/ \partial \phi$. The Coulomb gas charge 
$n_l$, corresponding to an elementary plaquette of the square lattice
$l$, is given by the directed sum (corresponding to a line integral)
over the four links $\langle ij\rangle$ making up the plaquette:~\cite{olsson}
\[
n_l\equiv \frac{T^{\rm CG}}{T}\sum_{\langle ij\rangle\in l}U'.\]
The correlation function $\hat{G}(k,t)$ is a key quantity and
is defined by 
\[ \hat{G}(k,t) \equiv \frac{1}{\Omega}\langle \hat{F}(k,t)
\hat{F}(-k ,0)\rangle ,  \]
where $\hat{F}(k, t)$ is the 1D Fourier transform
\[ \hat{F}(k,t)=\sum_mF_m(t) e^{ikm}, \]
$m$ labels the rows of the lattice, and finally
\[ F_m(t)=J\sum_{\langle ij\rangle \in m}U'[\phi_{ij}(t)] , \]
where the summation is over all the links making up the row $m$.
The Fourier transformation of the charge density correlation 
function $\hat{g}(k,t)$ is related to $\hat{G}(k,t)$ by
\begin{equation}
\hat{G}(k,t) = \left(
\frac{T}{T^{\rm CG}}\right)^2\frac{\hat{g}(k,t)}{k^2} .
\label{G}
\end{equation}
Linear-response theory then links
$\hat{g}(k,t)$ with the dielectric response function $1/\hat{\epsilon}(k,\omega)$ by~\cite{jonsson}
\begin{equation} \label{eq_Re_eps}
  {\rm
    Re}\left[\frac{1}{\hat{\epsilon}(k,\omega)}\right]=
    \frac{1}{\hat{\epsilon}(k,0)}+\frac{2\pi\omega
      T^{\rm CG}}{T^2}\int_0^\infty dt \sin \omega t\, \hat{G}(k,t) , 
    \label{reps}
\end{equation} 
        \begin{equation} \label{eq_Im_eps}
        {\rm Im}\left[ \frac{1}{\hat{\epsilon}(k,\omega)}\right] =
       -\frac{2\pi\omega T^{\rm CG}}{T^2}\int_0^\infty dt \cos \omega t
       \, \hat{G}(k,t) , 
       \label{ieps}
     \end{equation}
\noindent where
     \begin{equation}
       \frac{1}{\hat{\epsilon}(k,0)}=1-\frac{2\pi
       T^{\rm CG}}{T^2}\hat{G}(k,0)  .
     \label{eps}
     \end{equation}
The quantities $1/\hat{\epsilon}(0,\omega)$ and $\hat{G}(0,t)$ will be
of particular interest in the present investigation.

The thermodynamic KT transition is characterized by
\[\lim_{k\rightarrow 0} \frac{1}{\hat{\epsilon}(k,0)}=\frac{1}{\tilde{\epsilon}}>0 \]
below
the transition and
\[\lim_{k\rightarrow 0} \frac{1}{\hat{\epsilon}(k,0)}=0\]
above.
 Precisely at the transition $\lim_{k\rightarrow 0}
1/\hat{\epsilon}(k,0)T^{\rm CG}$ 
jumps from the universal value $1/\tilde{\epsilon}T^{\rm CG}=4$ to
zero.~\cite{nelson-kosterlitz,warren}
The equal-time correlations fall off like power laws with distance
below the
 transition and exponentially above.~\cite{minnhagenrev} For example,  the
correlation
 function $G(r,t=0)$ falls off like 
\begin{equation}
G(r,0)\propto r^{-\left(1/\tilde{\epsilon}T^{\rm CG}-2\right)}   
\label{gr}
\end{equation}
below the transition temperature.
The fact that the correlations decay algebraically with distance
reflects that 
the whole low-temperature phase is quasicritical. 

As explained in the previous section one motivation for the present
paper is the question
of the generality of the MP form for the dynamical response, which is given by~\cite{minnhagenrev}
\begin{equation} \label{eq_MP_Re}
  {\rm Re}\left[\frac{1}{\hat{\epsilon}(k=0,\omega)}-\frac{1}{\hat{\epsilon}(0,0)} \right]=\frac{1}{\tilde{\epsilon}}
\frac{\omega}{\omega +\omega_0}  ,
\label{rp}
\end{equation}
\begin{equation} \label{eq_MP_Im}
{\rm Im}\left[\frac{1}{\hat{\epsilon}(k=0,\omega)}\right]=-
\frac{2}{\tilde{\epsilon}\pi}\frac{\omega\omega_0\ln\omega/\omega_0}
{\omega^2 -\omega_0^2}  .
\label{ip}
\end{equation}  
The characteristic frequency $\omega_0$ vanishes as the KT transition
is approached from above and below.~\cite{jonsson} The idea behind the MP form
is that it describes the response due to the bound pairs. Consequently,
it is expected to have the correct leading small-frequency behavior below
the KT transition whereas it can only be approximately correct above because
of the presence of free vortices which always dominates the response for small
enough frequencies and gives a Drude-like response in this limit.~\cite{jonsson}
In the present paper we focus on the low-temperature phase. In this case 
the leading small $\omega$ behavior of Eqs.~(\ref{rp}) and (\ref{ip})
 reflects a $1/t$ decay for large $t$ of the
function $\hat{G}(k=0,t)$.~\cite{minnhagen1}
One may also observe that Eq.~(\ref{eq_MP_Im}) leads to a logarithmic divergence of the
real part of the conductivity: $\sigma(\omega) \sim - \omega {\rm Im}[1/\hat{\epsilon}(k=0,
\omega)] \sim - \ln\omega$  for small $\omega$, which is compatible with standard scaling
argument by Fisher and Fisher, Fisher, and Huse in Ref.~\onlinecite{fisher}.~\cite{jonssonL}

The two features
$G(r,t=0)\propto r^{-[(1/\tilde{\epsilon}T^{\rm CG})-2]}$ and
$\hat{G}(k=0,t)=\int d^2rG(r,t)\propto 1/t$ can be turned into an argument for the
dynamical critical index $z$ in the following way:~\cite{minnhagen1}
We assume that $G(r,t)$ must be of the form 
\[
G(r,t)\propto \lambda^\alpha f(r/\lambda,t/\tau ,a/r,\tau_a/t)  ,
\]
where $\lambda$ is the correlation length or screening length which
diverges in the low-temperature phase, $\tau$ is the corresponding
diverging relaxation time so that
\[
\tau\propto \lambda^z  ,
\]
where $z$ is the dynamical exponent. In addition we have a short
distance scale $a$, i.e., the lattice constant or the size of a Coulomb gas 
particle and a nondiverging characteristic time scale $\tau_a$, i.e., 
$\tau_a \propto l^2/D$ where $D$ is a vortex or Coulomb particle 
diffusion constant and $l$ is some nondiverging length scale
like $l=a$ or $l=1/\sqrt{n}$ where $n$ is the density of Coulomb gas particles.
Let us choose $t=0$ and $r=\lambda$ so that
\[
G(r,0)\propto r^\alpha f(1,0,a/r,\infty )
\]
and make the {\em ad hoc} scaling assumption that 
\[
\lim_{r\rightarrow \infty}f(1,0,a/r,\infty )=f(1,0,0,\infty )= {\rm const}  ,
\]
where const$\neq 0$ and $\neq\pm\infty$. This requires
$\alpha=-1/\tilde{\epsilon}T^{\rm CG}+2$ since $G(r,0)\propto
r^{-[(1/\tilde{\epsilon}T^{\rm CG})-2]}$.
We then also have that
\[
\int d^2rG(r,t)=\lambda^{-(1/\tilde{\epsilon} T^{\rm CG})+2}\int
d^2r f(r/\lambda,t/\tau ,a/r,\tau_a/t)  .
\]
Now we choose $\lambda=t^\frac{1}{z}$ so that
\[
\int d^2r G(r,t)= t^{[-1/\tilde{\epsilon} T^{\rm CG}+2]/z}\int d^2r
f(r/t^{1/z},1,a/r,\tau_a/t)
\]
and assume that
\[
\lim_{t\rightarrow \infty}
f(r/t^{1/z},1,a/r,\tau_a/t)=f(0,1,a/r,0)=\tilde{f}(a/r)  , 
\]
where $\tilde{f}(x)$ is a well-behaved function so that
\[
\int d^2r G(r,t)\propto t^{[-(1/\tilde{\epsilon} T^{\rm CG})+2]/z}\int
d^2r \tilde{f}(a/r)
\]
for large $t$. This is consistent with $\int d^2r G(r,t)\propto 1/t$
provided
\begin{equation} \label{eq_zscaling}
z=\frac{1}{\tilde{\epsilon}T^{\rm CG}}-2  .
\label{z}
\end{equation}

The dynamical exponent $z$ given by Eq.~(\ref{z}) has been inferred
through simulations of the lattice Coulomb gas with Monte
Carlo dynamics.~\cite{weber} In the present paper we
conclude that the same is true for the $XY$ models both
with RSJ and TDGL dynamics.

It has been argued by Dorsey,~\cite{dorsey} using scaling analysis, that for a 2D superconductor
the exponent $a$ in the nonlinear $IV$ characteristics $V\propto I^a$
has the value $a=z+1$ precisely at the KT transition. It has
further been suggested by
Minnhagen~\cite{minnhagen1} that since the whole low-temperature phase
is quasicritical the same relation should apply throughout the low-temperature 
phase. This together with Eq.~(\ref{z}) leads to the
prediction
\begin{equation}
a=a_{\rm scale}=z+1=\frac{1}{\tilde{\epsilon}T^{\rm CG}}-1   .
\label{ascale}
\end{equation}
The nonlinear $IV$ exponent $a=a_{\rm scale}$ in Eq.~(\ref{ascale}) has
been inferred through simulations for the Coulomb gas model with
Langevin dynamics~\cite{holmlund} and the lattice Coulomb gas model
with Monte Carlo dynamics.~\cite{weber}

The response to an imposed current is for a 2D superconductor 
given by
the complex impedance $Z(\omega )$:~\cite{minnhagenrev,hapnel}
\[ {\bf E}(\omega )=Z(\omega ){\bf j}(\omega )  ,
\]
where ${\bf E}(\omega )$ is the frequency dependent electric field and ${\bf
  j}(\omega )$
is the  current density. Or equivalently for a quadratic sample
$V(\omega )= Z(\omega )I(\omega )$, where $V$ is the voltage across the
  superconductor in some direction and $I$ is
the total current in the same direction. The linear-response function
$Z^{-1}(\omega )$ is related to the Coulomb gas linear-response
function $1/\hat{\epsilon}(k=0, \omega )$ by
\begin{equation}
  Z^{-1}(\omega )\propto \frac{\rho_0}{i\omega\hat{\epsilon}(k=0, \omega )}  ,
\label{imp}
\end{equation}
where $\rho_0$ is the density of superconducting electrons which for
an $XY$ model is given by $J\langle U''\rangle$. This
means that the effect on the vortex fluctuations of an imposed
current is given by $1/\hat{\epsilon}(k=0, \omega )$. For
small $\omega$ this is the dominant contribution.

It is instructive to consider the linear response to an imposed
current directly in the the case of the $XY$ model with RSJ dynamics. Let us consider a
quadratic lattice and let $\langle ij\rangle_x$ be a link at position
${\bf r}$ parallel to the $x$ axis and denote the
difference in phase angle by $\phi_{ij}=\nabla_x\theta ({\bf  r})$;
when the coupling to the electromagnetic field is included $\phi_{ij}$ denotes the
gauge invariant phase difference. The supercurrent through the link at
time $t$ is $JU'[\nabla_x\theta ({\bf r}, t)]$ and the normal current is
proportional to $-\nabla_x\dot{\theta}({\bf r},t)$ where the dot denotes
the time derivative. Thus the total current
$i_x({\bf r}, t)$ through the link is
\begin{equation}
i_x({\bf r}, t)=-\nabla_x\dot{\theta}({\bf r},t) +J U'[\nabla_x\theta ({\bf
  r}, t)]
\label{ix}
\end{equation}
in some convenient unit system.
The voltage in the RSJ model is proportional to the normal current so
we can define the response function corresponding to the complex
impedance as $Z({\bf r}-{\bf r}', t-t')=\dot{P}({\bf r}-{\bf r}',
t-t')$,  where
\begin{equation}
  P({\bf r}-{\bf r}', t-t')=-\left.\frac{\partial \langle \nabla_x\theta ({\bf
      r},t)\rangle}{\partial i_x({\bf r}',t')}\right|_{i_x=0}.
  \label{P}
\end{equation}
It is shown in the appendix that the Fourier transform of $P$ is given by
\begin{equation}
\hat{P}({\bf k}, \omega )=\left[ i\omega
  +\frac{\rho_0}{\hat{\epsilon}({\bf k}, \omega )}\right]^{-1} ,
  \label{P3}
\end{equation}
where $\rho_0=J\langle U''\rangle$ so that
\begin{equation}
  \hat{Z}({\bf k}, \omega )=\left[1 +\frac{1}{i\omega
      }\frac{\rho_0}{\hat{\epsilon}({\bf k}, \omega )}\right]^{-1}  .
 \label{Z}
\end{equation}
This means that the response to a uniform time varying current is given by
 $Z(\omega)=\hat{Z}(0, \omega )$. Below the KT transition we
have
\[\lim_{\omega \rightarrow 0}\lim_{{\bf k}\rightarrow
  0}\frac{1}{\hat{\epsilon}({\bf k}, \omega )}=\infty
\]
so that the static response to
a uniform static current below the KT transition is
nonlinear. However, for any finite frequency the response is linear to
the lowest order. One also notes that in the limit of high frequency
$1/i\omega \hat{\epsilon}({\bf k}, \omega )$ vanishes and $\hat{Z}$ in
Eq.~(\ref{Z}) reduces to $Z(\infty )=1$, which means that the response 
in this limit is
given by the resistive shunt in the RSJ model. For smaller frequencies
the response is given by the vortex fluctuation
$Z(\omega )\propto i\omega
\hat{\epsilon}(0,\omega )/\rho_0$ as already stated in Eq.~(\ref{imp}).

\section{Dynamical Equations and Boundary Conditions} \label{sec_dyn}

Simulations by necessity involve lattices with a finite linear
dimension $L$ from which the results for the thermodynamic limit
$L\rightarrow \infty$ have to be extracted. This means that in
practice the choice
of boundary condition is essential.~\cite{foota} The most commonly used boundary
condition in order to extract the thermodynamic limit for
the $XY$ models is periodic boundary conditions (PBC)
imposed on the phase angles $\theta_i$. However, as discussed in
Ref.~\onlinecite{olsson}, the PBC for the phase angles leads to a nonperiodic
boundary condition for the vortex interaction. The boundary condition
for the phase angles which corresponds to a periodic vortex interaction
is instead the fluctuating twist boundary condition (FTBC).~\cite{olsson} The
dynamics we are investigating in the present paper are linked to the vortex
fluctuations and consequently the natural boundary condition is
PBC for the vortices. This is the commonly used boundary condition for
simulations of the lattice Coulomb gas with Monte Carlo dynamics~\cite{weber}
and the continuum Coulomb gas with Langevin dynamics.~\cite{holmlund}
Thus the important point in the present context is that {\em PBC
for the vortices} means  {\em FTBC for the phase
angles}. The FTBC for the phase angles has so far been used in connection with  Monte Carlo
simulations.~\cite{olsson} In the present paper we extend the use of
these boundary conditions to $XY$ models with RSJ and TDGL dynamics.~\cite{footb}
Of course the boundary condition should not matter in the limit
$L\rightarrow \infty$. However, we in the present paper find that by
using FTBC for the phase angles we are able to extract more
information from our finite $L$ simulations.

In this section, we briefly review the dynamical equations of motion
for RSJ and TDGL in the case of PBC for the phase angles. Then
we construct the equations of motion for FTBC
starting from total current conservation and the condition that 
the equations of motion should lead to the correct equilibrium
distribution.
We focus on the ordinary $XY$ model, which 
corresponds to $p=1$ case in the previous section, but the extension 
to a general $p$ is straightforward.

We begin with an $L\times L$ array of the resistively shunted junctions with PBC 
in both directions. In the RSJ dynamics of 2D $XY$ model the net current from site $i$ to site $j$ is written as 
the sum of the supercurrent, the normal resistive current, and the thermal noise current:
\[
i_{ij} = i_c \sin(\phi_{ij}=\theta_i - \theta_j) + \frac{V_{ij}}{r} + \Gamma_{ij},
\]
where $i_c \equiv 2e J/ \hbar$ is the critical current of the single junction, 
$V_{ij}$
is the potential difference across the junction,  $r$ is the 
shunt resistance,
and the phase angles $\theta_i$ are periodic in both directions 
($\theta_i = \theta_{i+L{\hat x}} = \theta_{i+L{\hat y}}$).
The thermal noise current $\Gamma_{ij}$ at temperature $T$ is required 
to satisfy
$\langle \Gamma_{ij}(t)\rangle =0$ and $\langle \Gamma_{ij}(t) \Gamma_{kl}(0) \rangle = 
(2k_BT/r)\delta(t)(\delta_{ik}\delta_{jl} - \delta_{il}\delta_{jk})$.
The current-conservation law at each site, together with the Josephson
 relation 
$d(\theta_i - \theta_j)/dt = 2eV_{ij}/\hbar$, allows us to write the
 equations of motion in the form
\begin{equation}\label{eq_rsj_pbc}
\dot\theta_i = -\sum_j G_{ij}{\sum_k}^{'} [\sin(\theta_j - \theta_k) + \eta_{jk} ],
\end{equation}
where the primed summation is over four nearest neighbors of $j$, $G_{ij}$ is the lattice 
Green function on the square lattice with PBC, $\eta_{jk}$ is the dimensionless thermal
noise current defined by $\eta_{jk} \equiv \Gamma_{jk}/i_c$, and
the unit of time is $\hbar/2eri_c$. 
The thermal noise current 
satisfies $\langle \eta_{ij}(t) \rangle = 0$ and
\begin{equation} \label{eq_rsj_eta}
\langle \eta_{ij}(t) \eta_{kl} (0)\rangle =
2T(\delta_{ik}\delta_{jl} -  \delta_{il}\delta_{jk})\delta(t) ,
\end{equation}
where $T$ is in units of $J/k_B$.

In the TDGL dynamics with PBC, on the other hand, the equations of motion are given by~\cite{houlrik}
\[
\hbar \frac{d \theta_i (t)}{dt} = -\Gamma \frac{\partial H}{\partial \theta_i} + \Gamma_i (t) ,
\]
where $\Gamma$ is a dimensionless constant which determines the time scale of relaxation,
$H \equiv -J\sum_{\langle ij\rangle}\cos(\theta_i - \theta_j)$ is 
the Hamiltonian of the usual $XY$ model, and $\theta_i$ is periodic in 
both directions.  The thermal noise term $\Gamma_i (t)$ is assumed to satisfy
$\langle \Gamma_i (t) \rangle = 0$ and $\langle \Gamma_i (t) \Gamma_j (0) \rangle 
= 2 \hbar \Gamma k_BT \delta_{ij}\delta(t)$.
After rescaling the time and the temperature in units of $\hbar /\Gamma J$ and $J/k_B$,
respectively, the equations of motion for TDGL dynamics are written as
\begin{equation} \label{eq_tdgl_pbc}
\dot{\theta}_i  = -{\sum_j}^{'}\sin(\theta_i - \theta_j) + \eta_i ,
\end{equation}
where the thermal noise term $\eta_i \equiv \Gamma_i/\Gamma J$ 
satisfies $\langle \eta_i (t) \rangle = 0$ and
\begin{equation} \label{eq_tdgl_eta}
\langle \eta_i (t) \eta_j (0) \rangle = 2T \delta_{ij}\delta(t).
\end{equation}
In numerical simulations for PBC, we use Eqs.~(\ref{eq_rsj_pbc}) and (\ref{eq_tdgl_pbc})
for RSJ and TDGL dynamics, respectively, with the corresponding thermal noises
satisfying Eqs.~(\ref{eq_rsj_eta}) and (\ref{eq_tdgl_eta}). 

Next we consider the fluctuating twist boundary condition FTBC. In
this case a variable ${\bf \Delta} \equiv (\Delta_x, \Delta_y)$
is introduced and the phase
difference $\phi_{ij}$ on the bond $(i,j)$ is changed into~\cite{olsson}
\begin{equation} \label{eq_phase_diff}
\theta_i - \theta_j - {\bf r}_{ij} \cdot {\bf \Delta} ,
\end{equation}
where ${\bf r}_{ij} \equiv {\bf r}_j - {\bf r}_i$ is a unit vector from site $i$ to $j$,  
and the phase angles are periodic:
$\theta_i = \theta_{i + L{\hat x}} = \theta_{i + L{\hat y}}$. 
In the study of equilibrium behaviors for FTBC using MC simulations, 
it is sufficient to know the Hamiltonian of the system~\cite{olsson}
\begin{equation} \label{eq_H}
H = -J \sum_{\langle i j\rangle} \cos(\theta_i - \theta_j - {\bf r}_{ij} \cdot {\bf \Delta}).
\end{equation}
In dynamical simulations, on the other hand,
we must also have equations of motion for the new variables $\Delta_x$ and $\Delta_y$
in addition to the equations of motion for phase variables $\theta_i$.

The physical situation we have in mind is a sample where no current
passes through the boundary. For the RSJ model, which has
local current conservation, this implies the total current conservation
condition $\int dr^2 {\bf i}({\bf r},t)=0$,
where ${\bf i}({\bf r},t) = [i_x({\bf r},t), i_y({\bf r},t)]$ is the total current density
at point ${\bf r}$ and the integral is over the whole sample. This condition can also
be expressed as  
\begin{equation} \label{eq_Vxy}
\frac{ V_{x} } {r} = -\frac{i_c}{L} \sum_{\langle ij\rangle_{x}} 
\sin(\theta_i - \theta_j - \Delta_x) -\eta_{\Delta_x}  
\end{equation}
(and the similar equation for the $y$ direction),
where the summation $\sum_{\langle i j \rangle_x}$ is over
all nearest-neighboring pairs in $x$ direction,
$V_x$ is the voltage drop over the sample, and 
$\eta_{\Delta_x}$ denotes the thermal noise current. 
This follows because the left-hand side is recognized as 
the normal current whereas the right-hand side is the
negative of the sum of the supercurrent and the noise current.
As discussed in connection with Eq.~(\ref{ix}) the voltage is
by the Josephson relation proportional to  ${\bf
    \nabla}\dot{\theta}({\bf r},t)$.
For the voltage across the sample this means that [see Eq.~(\ref{eq_phase_diff})]
\begin{equation} \label{eq_dot_delta2}
\dot{\Delta}_{x} = - \frac{2e}{\hbar L} V_x ,
\end{equation}
because the phase angles are by construction subject to periodic
boundary conditions. Thus from  Eqs.~(\ref{eq_Vxy}) and (\ref{eq_dot_delta2}), we obtain the 
equations of motion for the twist variables:
\begin{eqnarray}
\frac{d{\Delta}_x}{dt} &=& \frac{1}{L^2} \sum_{\langle ij\rangle_x} 
                  \sin(\theta_i - \theta_j - \Delta_x) +\eta_{\Delta_x}   \label{eq_dot_deltax} , \\
\frac{d {\Delta}_y }{dt} &=& \frac{1}{L^2} \sum_{\langle ij\rangle_y} 
                  \sin(\theta_i - \theta_j - \Delta_y) +\eta_{\Delta_y}  \label{eq_dot_deltay}  ,
\end{eqnarray}
where we have again written $t$ in units of 
$\hbar /2eri_c$. Next a noise correlation consistent with the
equilibrium condition has to be found. To this end we make the ansatz
of a standard white-noise correlation 
$\langle \eta_{\Delta_x}(t) \eta_{\Delta_x} (0)  \rangle =
\langle \eta_{\Delta_y}(t) \eta_{\Delta_y} (0)  \rangle =
\sigma_\Delta^2 \delta(t)$ and determine the appropriate
$\sigma_{\Delta}^2$ in the following way: 
The equations of motion for the phase variables in FTBC are written as 
\begin{equation}\label{eq_theta}
\dot\theta_i = h_i - \sum_j {\sum_k}^{'} G_{ij} \eta_{jk} ,
\end{equation}
with 
\begin{equation}
h_i \equiv -\sum_j G_{ij}{\sum_k}^{'} \sin(\theta_j - \theta_k - \Delta_{jk}) \label{hi}
\end{equation}
and 
\[
\langle \eta_{ij}(t) \eta_{kl} (0)\rangle =
\sigma^2(\delta_{ik}\delta_{jl} -  \delta_{il}\delta_{jk})\delta(t) ,
\]
where $\sigma^2 = 2T$ [see Eq.~(\ref{eq_rsj_eta})].
From the full equations of motion for RSJ model in FTBC [Eqs.~(\ref{eq_dot_deltax}) -- (\ref{eq_theta})],
we arrive at the Fokker-Planck equation:~\cite{fokker}
\begin{eqnarray}
\frac{\partial W}{\partial t} &=& - \sum_i \frac{\partial}{\partial \theta_i} (h_i W) -
\frac{\partial}{\partial \Delta_x} (h_x W) - \frac{\partial}{\partial \Delta_y} (h_y W)  \nonumber \\
& &+ \frac{1}{2}\sigma^2\sum_{i,j} G_{ij}\frac{\partial^2 W}{\partial \theta_i \partial \theta_j}
+\frac{1}{2}\sigma_\Delta^2 \left( \frac{\partial^2 W}{\partial \Delta_x^2}
 + \frac{\partial^2 W}{\partial \Delta_y^2} \right), \nonumber
\end{eqnarray}
where $W = W(\{\theta_i\}, \Delta_x, \Delta_y ; t)$ is the probability distribution function
and 
\[
h_x \equiv \frac{1}{L^2}\sum_{\langle i j\rangle_{x} }\sin(\theta_i -\theta_j - \Delta_x),
\]
and the similar equation for $h_y$.
The stationary solution, which satisfies $\partial W/\partial t = 0$,
is of the correct form
$W = e^{-\beta H}$
with the Hamiltonian given by  Eq.~(\ref{eq_H}) provided 
\begin{eqnarray}
\frac{\beta\sigma^2}{2} &=& \beta T = 1, \label{eq_beta} \\
\frac{\beta\sigma_\Delta^2} {2} &=& \frac{1}{L^2}\label{eq_sigma}  ,
\end{eqnarray}
and consequently $\sigma_\Delta^2=2T/L^2$.

The equation of motion for the twist variables are hence of the
Langevin form
\begin{equation} \label{eq_dot_delta}
\dot{\bf \Delta } = -\Gamma_{\Delta} 
\frac{\partial H}{\partial {\bf \Delta}} + \eta_{\bf \Delta}
\end{equation}
with $\Gamma_\Delta=1/L^2$ and $\langle \eta_{\Delta_x}(t) \eta_{\Delta_x} (0)  \rangle =
\langle \eta_{\Delta_y}(t) \eta_{\Delta_y} (0)  \rangle = 
(2T/L^2) \delta (t)$.

In the TDGL model the total current conservation condition
can still be imposed whereas the local current conservation condition
is relaxed. Thus Eqs.~(\ref{eq_dot_deltax}) and (\ref{eq_dot_deltay})
remain unaltered 
whereas the equations for the phase angles are
simplified to [compare Eqs.~(\ref{eq_tdgl_pbc}) and (\ref{eq_phase_diff})]
\begin{equation} \label{eq_tdgl_tbc}
\dot{\theta}_i  = -{\sum_j}^{'}\sin(\theta_i - \theta_j - {\bf r}_{ij} \cdot {\bf \Delta}) + \eta_i  ,
\end{equation}
where we have used the dimensionless time $t$ by introducing the time unit of $\hbar/\Gamma J$ 
as in Eq.~(\ref{eq_tdgl_pbc}). Just as for the RSJ case one finds that 
$\Gamma_\Delta=1/L^2$ and that the noise correlation
$\langle \eta_{\Delta_x}(t) \eta_{\Delta_x} (0)  \rangle =
\langle \eta_{\Delta_y}(t) \eta_{\Delta_y} (0)  \rangle =
(2T/L^2) \delta(t)$ leads to the correct equilibrium. To some extent
the TDGL dynamics may be viewed as a simplified version
of the RSJ dynamics where the total current conservation is kept but
the local current conservation is relaxed. Thus from this point of
view it is perhaps not surprising that the two models (as we will see)
have the same generic vortex dynamics.

 The twist variable ${\bf \Delta}$ plays an important role in
our analysis of the vortex dynamics and there exists a rather direct
connection between the twist and the vortices:  
The electric field ${\bf E}(t)$ due to the vortex current
density ${\bf j}_v$  is perpendicular and is,
 as a consequence of the Josephson relation, given by~\cite{weber}
\[
E = \frac{h}{2e} \langle j_v(t) \rangle .
\]
The connection between $\langle {\bf j}_v(t) \rangle$ and
$\dot{\bf \Delta}$ is discussed in Ref.~\onlinecite{olsson}; when
a vortex moves across the sample then the twist variable changes by $2\pi /L$.
In other words, if the time $t_0$ is associated with the movement of a
vortex across the sample, then we get
$\dot{\bf \Delta} = 2\pi/Lt_0=2\pi \langle {\bf v}\rangle /L^2$ where $|\langle{\bf v}\rangle| = L/t_0$ is the vortex velocity. 
If there are $N_v$ moving vortices,
then we obtain $\dot{\bf \Delta} = 2\pi (N_v /L^2) \langle{\bf v}\rangle = 
2\pi \langle {\bf j}_v \rangle$, which leads to the relation given by Eq.~(\ref{eq_dot_delta2}):
\[
\dot{\Delta}_x =  -\frac{2e}{\hbar L} V_x ,
\]
where $V_x$ is the voltage drop across the sample in $x$ direction 
(we obtain the similar equation for $\Delta_y$).

So far we have considered the situation when the total current in
the sample is zero, which corresponds to no current passing over
the boundary. Let us now consider the case when the total current
is a constant dc current $I_d$ in the $x$ direction. By following
the steps from Eq.~(\ref{eq_Vxy}) to Eqs.~(\ref{eq_dot_deltax})
 and (\ref{eq_dot_deltay}) one obtains
the modified equations of motion for the twist variable $\Delta$   
\begin{eqnarray}
\frac{d{\Delta}_x}{dt} &=& \frac{1}{L^2} \sum_{\langle ij\rangle_x}
                  \sin(\theta_i - \theta_j - \Delta_x) +\eta_{\Delta_x} - i_d \label{eq_Deltax_Id}  \\
\frac{d {\Delta}_y }{dt} &=& \frac{1}{L^2} \sum_{\langle ij\rangle_y}
                  \sin(\theta_i - \theta_j - \Delta_y) +\eta_{\Delta_y}   \label{eq_Deltay_Id} 
\end{eqnarray}
with $i_d=I_d/L$ in units of $i_c$.  
The voltage drop in the $x$ direction [see Eq.~(\ref{eq_dot_delta2})] is given by 
\begin{equation} \label{eq_Vx}
V_x  = -L \dot{\Delta}_x
\end{equation}
with $V_x$ in units of $ri_c$ for RSJ and in units of $\Gamma J/2e$ for TDGL, respectively. Thus the equations of motion in the presence of an
externally imposed dc current $I_d$ in the $x$ direction 
are given by Eqs.~(\ref{eq_theta}), (\ref{eq_Deltax_Id}), and (\ref{eq_Deltay_Id})
for RSJ and by Eqs.~(\ref{eq_tdgl_tbc}), (\ref{eq_Deltax_Id}), and (\ref{eq_Deltay_Id}) for TDGL.

An alternative and commonly used method in numerical simulations
of the current-driven $XY$ model 
is to impose uniform currents through the boundary in one direction.
This requires an open boundary condition for the phase angles
in the direction of the applied current and 
the periodic boundary condition can only be kept 
in the perpendicular direction.~\cite{mon}
This means that an open boundary is explicitly introduced. One advantage with the present 
method is that the periodic boundary conditions on the phase angles 
are kept and no explicit boundary is introduced.
In the following two sections we present the results obtained from the
dynamical equations described in the present section both for the PBC 
and the FTBC.   

\section{Simulation Results} \label{sec_result}

In this section, we present simulation results for the TDGL and
RSJ dynamics with periodic boundary conditions PBC and the fluctuating twist boundary
conditions FTBC. For PBC, we use Eqs.~(\ref{eq_rsj_pbc}) and
(\ref{eq_rsj_eta}) in the RSJ case and in the TDGL case
Eqs.~(\ref{eq_tdgl_pbc}) and  (\ref{eq_tdgl_eta}). For FTBC, we use
Eqs.~(\ref{eq_dot_deltax}) -- (\ref{eq_theta}) for RSJ, and
Eqs.~(\ref{eq_dot_deltax}), (\ref{eq_dot_deltay}), and (\ref{eq_tdgl_tbc})
for TDGL.

We integrate the equations of motion by discretizing time into small
steps $\Delta t$.
At each step the appropriate random noise, generated from a uniform distribution, 
is introduced with $\langle \eta_{ij}(t)^2 \rangle = 2T/\Delta t$ for RSJ and
$\langle \eta_i(t)^2 \rangle = 2T/\Delta t$ for TDGL [see 
Eqs.~(\ref{eq_rsj_eta}) and (\ref{eq_tdgl_eta})]. 
We want to integrate to as long
times as possible.~\cite{footc} On the other hand the larger $\Delta t$ we choose
the larger is the error introduced by the discretization.
In order to get a handle of the choice for $\Delta t$ we use the following identity: 
Let us introduce a local variable $a_k$
on one particular site $k$. The Hamiltonian 
of the system is then
\[
H = \sum_{\langle ij\rangle} U(\theta_i - \theta_j + a_i - a_j)
\]
with $a_i \neq 0$ for $i=k$ and $a_i = 0$ otherwise,
and the partition function  is given by
\[
Z = \int\prod_i d\theta_i \exp(-\beta H)
\]
with the inverse temperature $\beta \equiv 1/T$. After a simple change of variable
$\theta_i + a_i \rightarrow \theta_i$, we find that $Z$
does in fact not depend on $a_k$ and thus 
\[
\frac{\partial^2 \ln Z}{\partial a_k^2} = 0 ,
\]
from which we conclude that
\[
 4 \langle U''\rangle = \frac{1}{T}\left\langle\left[{\sum_j}^{'} U'(\theta_k - \theta_j)\right]^2\right\rangle , \
\]
and thus 
\begin{equation} \label{eq_T}
T=\tilde{T}
\end{equation}
provided we have defined $\tilde{T}$ by the local correlations
\begin{equation}  \label{eq_Teff}
\tilde{T} \equiv \frac{\left\langle\left[{\sum_j}^{'} U'(\theta_k - \theta_j)\right]^2\right\rangle}
 {4 \langle U''\rangle} ,
\end{equation}
where the summation is over four nearest neighbors (denoted by $j$) of site $k$.
The point is now that for a finite $\Delta t$ one finds that
$\tilde{T}>T$. 
In the present simulations we use the time step 
$\Delta t$ = 0.01 for TDGL and $\Delta t$ = 0.05 
for RSJ. These choices make
$\tilde{T}$ differ from $T$ by less than $3\%$.

The fact that $\tilde{T}>T$ for a finite time step suggests that the
effect of the finite time step to some extent is equivalent to an
increased temperature. We have tried to take this into account when
analyzing quantities related to $1/\hat{\epsilon}$ by noting that
for FTBC one has $1/\hat{\epsilon}(0,0)=0$,~\cite{olsson} which means that [compare
Eq.~(\ref{eps})]
\[ T=\frac{\hat{G}(0,0)}{\langle U''\rangle} .  \]
Thus we can estimate an effective temperature by $T^{\rm
  eff}=\hat{G}(0,0)/\langle U''\rangle$. For example, for $T=0.80$ and
  the time step $\Delta t=0.05$ we for RSJ get $T^{\rm eff}\approx 0.82$
  whereas we for TDGL and the time step $\Delta t=0.01$ get $T^{\rm
  eff}\approx 0.803$.
\subsection{Dynamical response functions with periodic
 boundary conditions}
\begin{figure}
\centerline{\epsfxsize=8.0cm \epsfbox{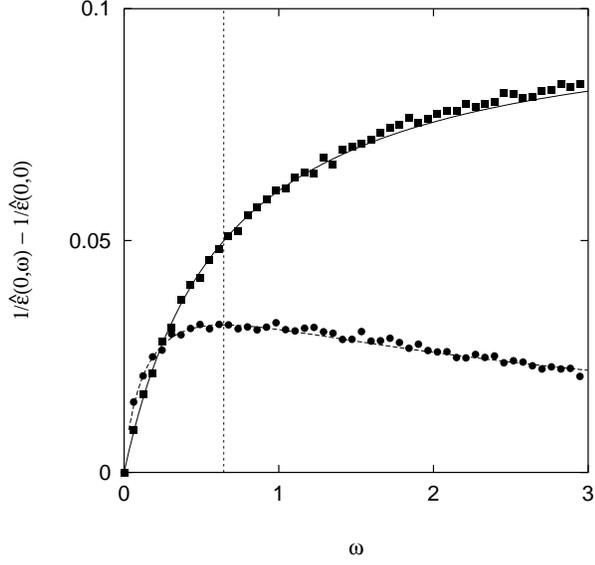}}
\vskip  -2cm
\caption{The dynamical response function $1/\hat{\epsilon}(0,\omega)$ 
of the 2D $XY$ model with RSJ dynamics 
at $T=0.85$ for a $64 \times 64$ lattice with periodic boundary conditions. 
[The frequency $\omega$ is in units of $2eri_c /\hbar$ (see text).]
The filled
squares and circles correspond to the real part and the absolute
value of the imaginary
part of the dynamical response function, respectively.
The full curve is obtained by fitting to the
real part of the MP form response function in Eq.~(\ref{eq_MP_Re}) 
and the broken curve is the imaginary part 
Eq.~(\ref{eq_MP_Im}) using the same values of the fitting parameters as 
for the full curve.
The vertical broken line corresponds to the $\omega$ for which the peak ratio 
$|{\rm Im}[1/\hat{\epsilon}(0,\omega)]|/{\rm Re}[1/\hat{\epsilon}(0,\omega) - 1/\hat{\epsilon}(0,0)]$ 
is $2/\pi$. At this $\omega$ the absolute
value of the imaginary part
should, accordingly to the MP form, have a maximum.
}
\label{fig1}
\end{figure}

We will first consider the vortex dynamics as reflected in the complex
dielectric
function given by Eqs.~(\ref{eq_Re_eps}) and (\ref{eq_Im_eps}).
It has so far been established that the MP form
Eqs.~(\ref{eq_MP_Re}) and (\ref{eq_MP_Im}) 
gives a good representation of the experimental data,~\cite{theron} as
well as the simulation data for the TDGL dynamics of the $XY$ model on a square lattice with 
$p=2$ and on the triangular lattice with $p=1$,~\cite{jonsson} and the 2D
Coulomb gas model.~\cite{holmlund}
In the present investigation we find that the same is true for the $XY$ model with RSJ dynamics. 
This is illustrated in Fig.~\ref{fig1} which shows the real
and imaginary parts of $1/\hat{\epsilon}({\rm k}=0, \omega )-
1/\hat{\epsilon}(0, 0)$
with RSJ dynamics for $T=0.85$. The full line in the figure has been
obtained from a least-square fit to
the MP form of the real part in Eq.~(\ref{eq_MP_Re})
with two free parameters
(${\tilde \epsilon}$ and $\omega_0$), and the broken line has been
obtained
by using the same values of the parameters in
Eq.~(\ref{eq_MP_Im}) (the frequency range in Fig.~\ref{fig1} corresponds to
$0.08 < \omega/\omega_0 < 4.7 $). 
The MP form has the characteristic feature that the 
ratio is $|{\rm Im}[1/\hat{\epsilon}(0,\omega )]|
/{\rm Re}[1/\hat{\epsilon}(0, \omega ) - 1/\hat{\epsilon}(0,0)]=2/\pi$ 
at the frequency where the imaginary part has its maximum. One sees
directly in Fig.~\ref{fig1} (i.e., without any curve fitting) that the
dotted vertical line is close
to this maximum and it is hence easy to verify that the ratio is
indeed
close to $2/\pi$.
In short, our present simulations of the complex dielectric
function confirm that 
the RSJ dynamics
is well described by the MP form at temperatures below as well as
somewhat above the critical temperature in agreement with what was found earlier for the TDGL dynamics in Ref.~\onlinecite{jonsson}.~\cite{footca} 

As pointed out in connection with Eqs.~(\ref{rp}) and (\ref{ip}), the
leading small $\omega$ dependence of the MP form
\[ {\rm Re}\left[
  \frac{1}{\hat{\epsilon}(0,\omega)}-\frac{1}{\hat{\epsilon}(0,0)}\right] \propto \omega\]
  and
  \[ {\rm Im}\left[\frac{1}{\hat{\epsilon}(0,\omega)}\right]\propto
      \omega\ln \omega \]
reflects that $\hat{G}({\bf k}=0,t)\propto 1/t$ for large $t$. More
precisely, since      
\[
\hat{G}(0,t) = \frac{T^2}{\pi^2T^{\rm CG}}\int_0^\infty \frac{\sin\omega t}{\omega}
{\rm Re}\left[ \frac{1}{\hat{\epsilon}(0,\omega)} 
- \frac{1}{\hat{\epsilon}(0,0)} \right]  d\omega,
\]
we find for the MP form
\begin{equation} \label{eq_GMP}
\hat{G}^{\rm MP} (0,t) = \frac{T^2}{\pi^2\tilde\epsilon T^{\rm CG}}
\left[ {\rm Ci}(\omega_0 t)\sin\omega_0 t - {\rm si}(\omega_0 t) \cos\omega_0 t \right],
\end{equation}
where the cosine and the sine integrals are defined by
${\rm Ci}(x) \equiv -\int_x^\infty dt \cos t/t$ and
${\rm si}(x) \equiv -\int_x^\infty dt \sin t/t$, respectively.
In the limit of $\omega_0 t \rightarrow \infty$, Eq.~(\ref{eq_GMP}) reduces to
\[
\hat{G}^{\rm MP} \approx \frac{T^2}{\pi^2\tilde\epsilon T^{\rm CG}} \frac{1}{\omega_0 t}.
\]
This $1/t$ tail in the vortex correlations has been verified in
Ref.~\onlinecite{minnhagen1} for TDGL dynamics and in Ref.~\onlinecite{holmlund} for
the Coulomb gas model. We will here verify the same result for the
RSJ dynamics.

\begin{figure}
\centerline{\epsfxsize=8.0cm \epsfbox{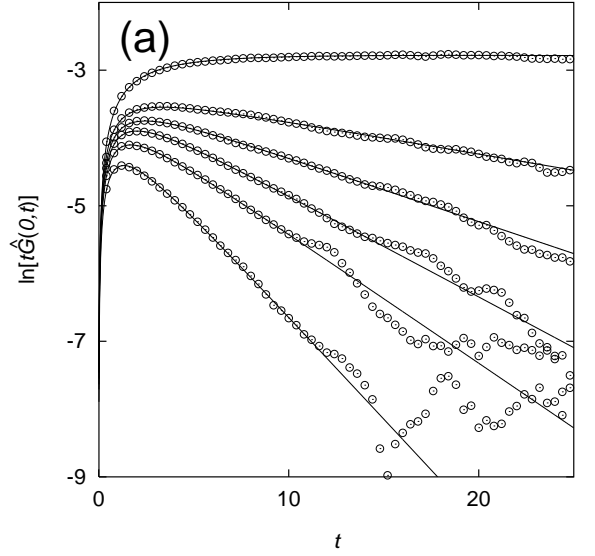}}
\vskip  -3cm
\centerline{\epsfxsize=8.0cm \epsfbox{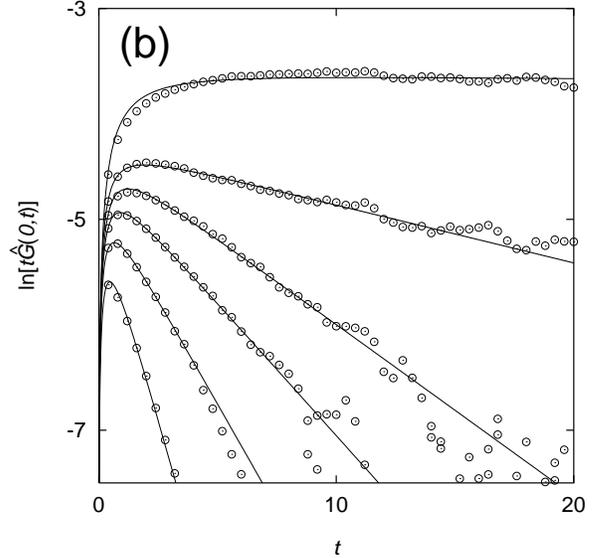}}
\vskip  -2cm
\caption{The time correlation function $\ln[t\hat{G}(0,t)]$ versus time $t$
at $T=0.85$ for various system sizes [$L=6,8,10,12,16$, and 64 from
bottom to top] in case of (a) RSJ and (b) TDGL dynamics.
The full curves have been obtained by fitting to the 
modified-MP (MMP) form Eq.~(\ref{eq_GMMP}). The figure shows that that the relaxation
time $\tau_G$ in the MMP form diverges as the system size is
increased and that $\hat{G}(0, t)\propto 1/t$ for large $t$ in the
thermodynamic limit.
}
\label{fig2}
\end{figure}

By necessity, the finite lattice sizes used in the simulations
introduce a finite relaxation time $\tau_G$ at large $t$ for the zero-$k$ mode. By
studying the lattice size dependence of $\hat{G}(0,t)$ we have found
that this finite size induced relaxation changes the large-$t$ decay from $1/t$
to $(1/t)\exp(-t/\tau_G)$. In fact we have found that $\hat{G}(0,t)$ for
finite lattices to a good approximation is of a modified-MP form (MMP):
\begin{equation} \label{eq_GMMP}
\hat{G}^{\rm MMP} \equiv \hat{G}^{\rm MP} \exp(-t/\tau_G) .
\end{equation}
Figure~\ref{fig2} shows $\ln[t\hat{G}(0,t)]$ as a function of time for the system
sizes $L=6,8,10,12,16$, and 64 in case of (a) RSJ and (b) TDGL dynamics at $T=0.85$.
The full drawn curves are least-square fits to Eq.~(\ref{eq_GMMP}). As
is apparent from Fig.~\ref{fig2}, $t\hat{G}$ approaches a constant
for large lattice sizes verifying that $\hat{G}$ indeed goes as $1/t$
for large $t$ both for RSJ and TDGL dynamics.

\begin{figure}
\centerline{\epsfxsize=8.0cm \epsfbox{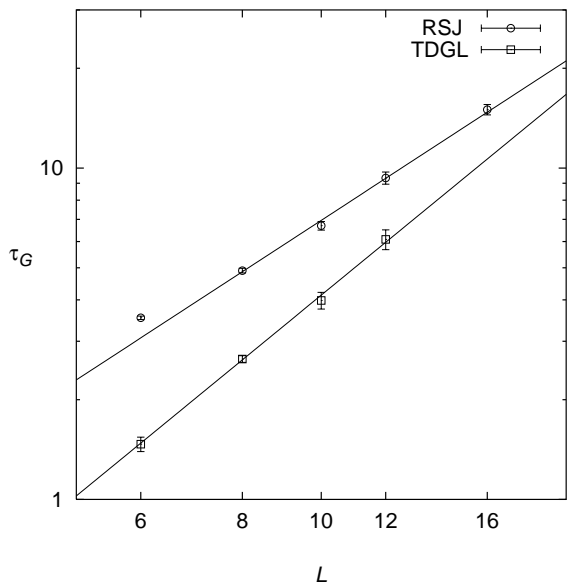}}
\vskip  -2cm
\caption{
The relaxation time $\tau_G$ for the time correlation function $\hat{G}(0,t)$
at $T=0.85$ for RSJ (squares) and TDGL (circles) dynamics.
The data points have been obtained from least-square fits to the MMP
form $\hat{G}^{\rm MMP}$ given by Eq.~(\ref{eq_GMMP}) as shown in Fig.~\ref{fig2}.
As the system size $L$ is increased $\tau_G$ diverges. However, the exponent $z$ 
defined by $\tau_G \sim L^{z}$ appears to have {\em different} values for
the two types of dynamics.
The lines are obtained from least-square fits using data points for $L=8$, 10, 12, and 16 
in the RSJ case and $L=6$, 8, 10, and 12 in the TDGL case, giving  $z \approx 1.6$ and 
$z \approx 2.0$ for RSJ and TDGL, respectively. 
}
\label{fig3}
\end{figure}

The fits to the MMP form (full drawn curves in Fig.~\ref{fig2}) show
that $\ln t\hat{G}(0,t)$ goes as $-t/\tau_G$ for large $t$. In
Fig.~\ref{fig3} we have plotted $\tau_G$ [determined by the fit to
Eq.~(\ref{eq_GMMP})] as a function of lattice size $L$ in a log-log
scale. From finite-size scaling we expect that in the low-temperature phase 
$\tau_G$ diverges as
$\tau_G\propto L^z$ for large $L$ where $z$ is the dynamical critical
exponent.
This behavior corresponds to straight lines in
Fig.~\ref{fig3} and the full straight lines in the figure suggest
that the asymptotic scaling is reached already for relatively small $L$. Assuming
that this is the case, we find from the slopes of the lines that for $T=0.85$
$z\approx 1.6$ in case of RSJ and $z\approx 2$ for TDGL. Thus the $z$ values
in case of PBC are {\em different} for the RSJ and the TDGL dynamics. 
This difference between RSJ and TDGL in case of periodic boundary
conditions was also found by Tiesinga {\em et al}.
in Ref.~\onlinecite{tiesinga}, where in the temperature interval $T\in [1.1,
1.3]$ $z\approx 2$ for TDGL and $z\approx 0.9$ for RSJ;
the authors concluded that the TDGL somewhat unexpectedly
describes the experiments on Josephson junction arrays by Shaw {\em et al.}~\cite{shaw} 
better than the RSJ model. The conclusion we arrive at
is different since we find that for FTBC the equivalence between RSJ and TDGL is
restored. The apparent difference in case of PBC appears to be a boundary effect.~\cite{footd}
We believe
that the physical situation in Ref.~\onlinecite{shaw} and most other
common experimental situations are in fact better described by the
FTBC. Of course, for large enough system sizes,
intensive physical quantities do not depend on the explicit
choice of boundary condition. But the point here is that, because the relaxation of
the zero-$k$ mode is described by a relaxation time $\tau_G$ which diverges
for infinite systems, the exponent $z$,
which describes how this divergence is approached, appears to be sensitive to the
choice of boundary condition.~\cite{footd}

We also note that for $T=0.90$ we find $z\approx 1.6$ in case of
RSJ with PBC. This suggests that $z$ for PBC approaches a value less than 2 as
the KT transition is approached from below, although the numerical accuracy
may be insufficient to make a firm conclusion.

\subsection{Dynamics for the fluctuating twist boundary conditions}
\begin{figure}
\centerline{\epsfxsize=8.0cm \epsfbox{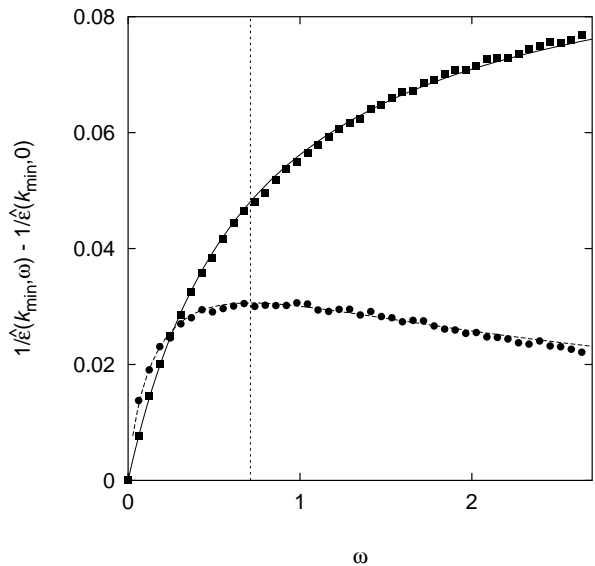}}
\vskip  -2cm
\caption{The dynamical response function $1/\hat{\epsilon}({\bf k},\omega)$ 
  at ${\bf k} = {\bf k}_{\rm min} \equiv (0, 2\pi/L)$ for a $L=64$ array at $T=0.85$  for RSJ 
dynamics with FTBC ($\omega$ is in units of $2eri_c /\hbar$).
The filled squares and circles correspond to the real part 
and (the absolute value of) the imaginary part of
 $1/\hat{\epsilon}({\bf k},\omega)$. The full curve
is obtained from a least-square fit to Eq.~(\ref{eq_MP_Re})
with two free parameters $\omega_0$ and $\tilde{\epsilon}$ and
the broken curve is obtained from Eq.~(\ref{eq_MP_Im}) using these parameter values.
The vertical broken line is given by the condition
that the peak ratio 
$|{\rm Im}[1/\hat{\epsilon}(k,\omega)]|/{\rm
 Re}[1/\hat{\epsilon}(k,\omega) - 1/\hat{\epsilon}(k,0)]=2/\pi$
and at this value of $\omega$, the absolute value of the imaginary part
should, according to the MP form, have a maximum.
}
\label{fig4}
\end{figure}

\begin{figure}
\centerline{\epsfxsize=8.0cm \epsfbox{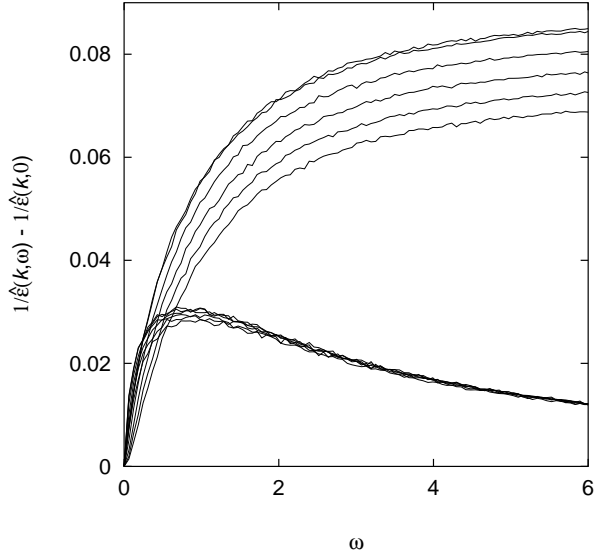}}
\vskip  -2cm
\caption{The dynamical response function $1/\hat{\epsilon}({\bf k},\omega)$ 
at finite ${\bf k}$ for a $L=64$ array at $T=0.85$  for RSJ dynamics with FTBC
($\omega$ is in units of $2eri_c /\hbar$).
The real and imaginary parts are obtained using the wave vectors
${\bf k} = (0, k_y = 2\pi n_y /L)$
with $n_y$ = 1, 2, 4, 6, 8, and 10 (from top to bottom). The imaginary part
depends very little on the value of $k$ in the frequency
interval around the maximum, in contrast to the real part which
increases with decreasing $k$. 
}
\label{fig5}
\end{figure}

\begin{figure}
\centerline{\epsfxsize=8.0cm \epsfbox{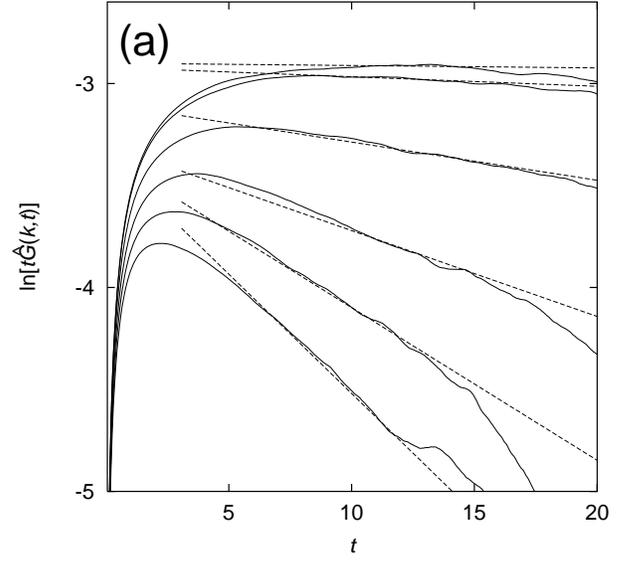}}
\vskip -3cm
\centerline{\epsfxsize=8.0cm \epsfbox{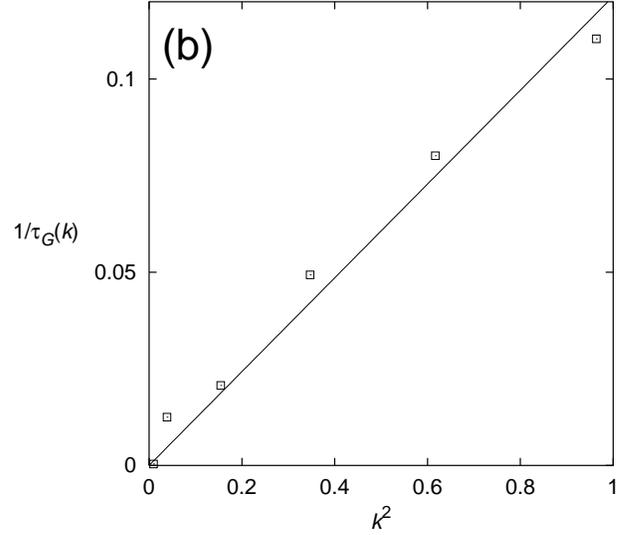}}
\vskip  -2cm
\caption{ (a) The time correlation function $\ln[t\hat{G}({\bf k},t)]$ versus time $t$
  at $T=0.85$ for RSJ dynamics with FTBC.
The wave vectors are ${\bf k} = (0, k_y = 2\pi n_y /L)$
with $n_y$ = 1, 2, 4, 6, 8, and 10 (from top to bottom) and the array size is $L=64$.
As ${\bf k} \rightarrow 0$, $\hat{G}({\bf k},t)$ approaches
 $\hat{G}({\bf k},t) \rightarrow 1/t$ for large values of $t$. At nonzero value
of $k$, $\hat{G}(k,t)$ exhibits exponential decay $\hat{G}(k,t) \sim 
\exp[-t/\tau_G(k)]/t$ in the long-time limit. The broken lines are plotted with
the $\tau_G(k)$ values corresponding to the straight line in (b), where we show $\tau_G(k)$ 
versus $k^2$. In (b), the squares have been obtained from the least-square fit of $\hat{G}(k,t)$
to the exponential decay form, and the full straight line is the result of the least-square
fit to  $\tau_G(k)\propto k^2$. }
\label{fig6}
\end{figure}

In case of FTBC the static dielectric function function
$1/\hat{\epsilon}({\bf k},0 )$ is identically zero for ${\rm k}=0$,
whereas $\lim_{{\bf k}\rightarrow 0}1/\hat{\epsilon}({\bf k},0)\neq 0$
below the KT transition.~\cite{olsson} In Ref.~\onlinecite{holmlund} it
was shown that for the Coulomb gas model with Langevin dynamics the
function $1/\hat{\epsilon}({\bf k},\omega)$ for small ${\bf k}$ is to good
approximation given by the MP form. Since, as explained above in
Sec.~\ref{sec_dyn}, PBC for
the vortices (as in Ref.~\onlinecite{holmlund}) corresponds to FTBC for the
$XY$ model we also expect to find the MP form for small $k$ in the
present case. This is illustrated in
Fig.~\ref{fig4} which shows the real and imaginary parts of
$1/\hat{\epsilon}({\bf k},\omega )$ for ${\bf k}=(0,2\pi/L)$ with
$L=64$ for the $XY$ model with RSJ dynamics. The full drawn and broken curves represent the
MP form just as in Fig.~\ref{fig1} and the dotted line shows that the peak
ratio is close to $2/\pi$.  Figure~\ref{fig5}
demonstrates that the imaginary part depends very little on the
$k$ value whereas the real part increases with decreasing $k$ for
fixed frequency. This behavior has also been found for the Coulomb gas model
with Langevin dynamics (compare Figs.~11 and 12 in
Ref.~\onlinecite{holmlund}). Figure~\ref{fig6} shows how the relaxation time
$\tau_G$ of $\hat{G}({\bf k},t)$ depends on $k$: $\hat{G}\propto
(e^{-t/\tau_G})/t$ for large $t$ and $\tau_G$ diverges as $k$
is decreased. In Ref.~\onlinecite{holmlund} it was found that
$\tau_G\propto k^{-2}$ for the Coulomb gas model with Langevin
dynamics. Our present convergence is not good enough for establishing this
result, but Fig.~\ref{fig6}(b) suggests that such a
behavior is also consistent with the present simulations
of the $XY$ model with RSJ dynamics.

\begin{figure}
\centerline{\epsfxsize=9.0cm \epsfbox{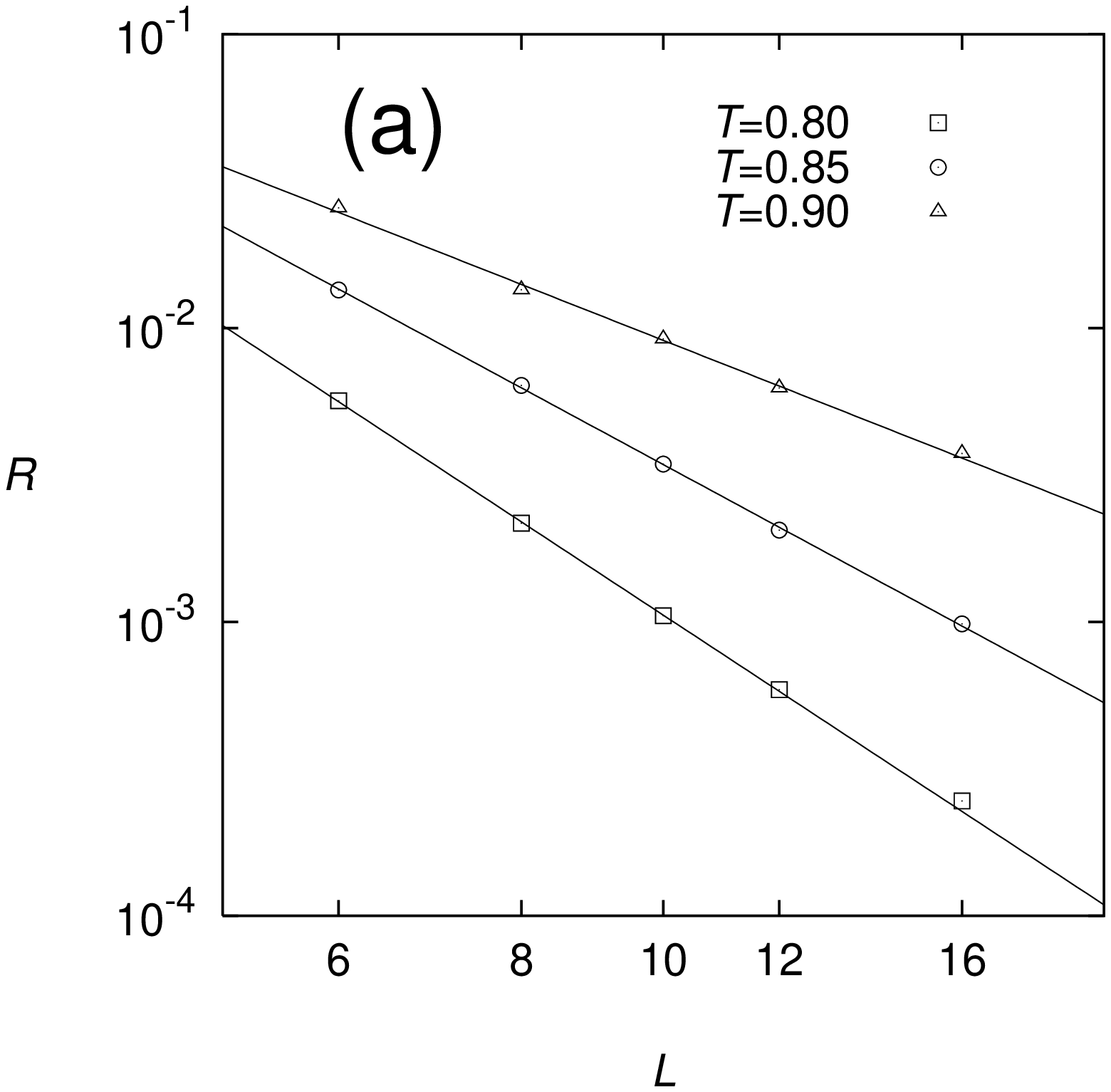}}
\vskip -4cm
\centerline{\epsfxsize=9.0cm \epsfbox{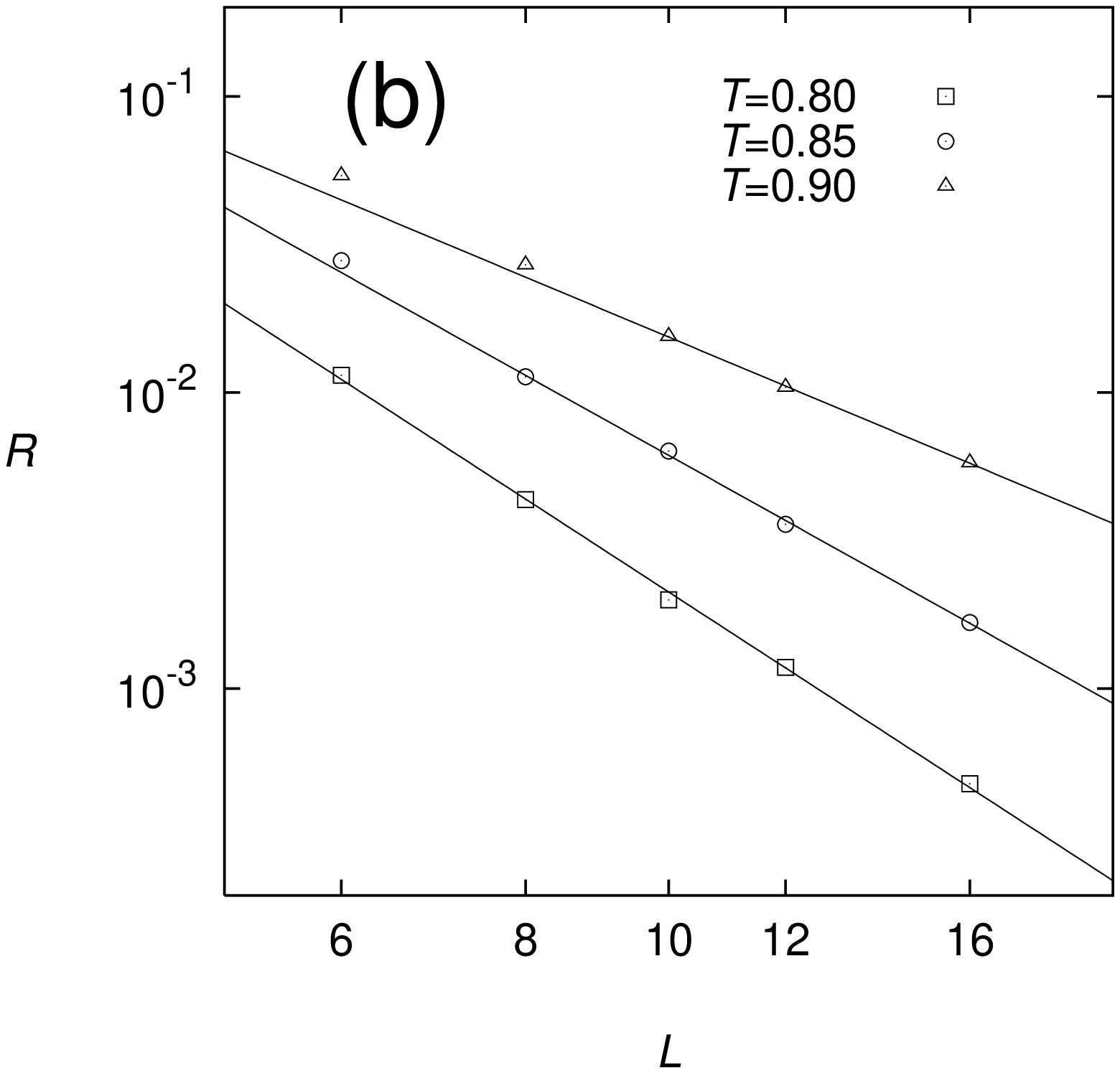}}
\vskip  -2cm
\caption{Resistance $R$ versus system size $L$ for (a) RSJ  and (b)
TDGL dynamics obtained from  Eq.~(\ref{eq_R}). The full lines
are obtained by fitting to the scaling form $R \sim L^z$ and from
these fits the values of $z$ are determined to be
$z$ = 3.3(1), 2.7(1), and 2.0(1) at $T=0.80$, 0.85, and 0.90 for the RSJ case,
and $z$ = 3.3(1), 2.8(1), and 2.1(1) at $T=0.80$, 0.85, and 0.90 for TDGL.
}
\label{fig7}
\end{figure}

Next we turn to the diverging relaxation time $\tau$ and the dynamical critical
exponent $z$ for the case of FTBC. We will use the fact that in the
low-temperature phase the resistance $R$ of a finite system is
proportional to $1/\tau$.~\cite{wallin} This follows because
of the Nyquist formula:~\cite{reif}
\begin{equation} \label{eq_Nyq}
R = \frac{1}{2k_B T} \int_{-\infty}^\infty dt \langle V(t) V(0) \rangle 
\end{equation}
which relates the resistance to the voltage fluctuations over the
sample and the fact that $V\propto (d/dt)\Delta \phi$ where
$\Delta \phi$ is the phase difference over the sample. Since $\Delta
\phi$ is dimensionless it follows that $R$ scales like $1/\tau$
where $\tau$ is the relevant characteristic time.~\cite{wallin} In the
low-temperature phase  $R$ vanishes in the limit of
large system sizes since $\tau$ diverges. Consequently the
finite-size scaling $R\propto 1/\tau\propto L^{-z}$ defines the value of the
dynamical critical exponent $z$ in the low-temperature phase. For the
$XY$ model with FTBC the phase difference over the
sample in one direction (let us choose the $x$ direction) is given by
$\Delta \phi=L\Delta_x$. It follows that $R$ can be expressed as    
\begin{equation} \label{eq_R}
R = \frac{L^2}{2 T } \frac{1}{\Theta} \langle [ \Delta_x(\Theta) - \Delta_x(0) ]^2 \rangle ,
\end{equation}
where 
$T$ is in units of $J/k_B$, $\Delta_x(t)$ is the twist variable in the $x$ direction at time $t$,
and $R$ is in units of the shunt resistance $r$ of a single
Josephson junction for the RSJ model
and $\Gamma J/2e i_c$ for TDGL model, respectively. 
Since Eqs.~(\ref{eq_Nyq}) and (\ref{eq_R}) are identical in the limit of large
$\Theta$, i.e., for $\Theta \gg \tau$,~\cite{reif}  we  
for practical reasons use Eq.~(\ref{eq_R}) in the present
simulations (we have used $\Theta = 16 000$ and $\Theta \gg \tau$). 
Figure~\ref{fig7}(a) shows the results for the $XY$ model with
RSJ dynamics for $T=$0.90, 0.85, and 0.80. The data are plotted as $\log R$
against $\log L$ and to good approximation fall on a straight line, 
whose slope gives an estimate of the critical exponent $z$,
and we obtain $z=$2.0, 2.7, and 3.3, respectively.
Figure~\ref{fig7}(b) shows the same features for the $XY$ model with
TDGL dynamics at the same three temperatures $T=$0.90, 0.85, and 0.80
and the estimated values of $z\approx$ 2.1, 2.8, and 3.3 are close to 
the ones obtained for the  RSJ dynamics.

Thus for the FTBC we find the same $z$ below the KT transition for RSJ 
and TDGL dynamics, which is in contrast to the PBC case where we found
different values of $z$ for each dynamics (compare the discussion
of Fig.~\ref{fig3} in Sec.~\ref{sec_result}). Furthermore for FTBC we find that $z$
apparently approaches 2 when the KT transition is approached from below ($T=0.90$
is very close to the KT transition temperature) for both dynamics; 
this did not seem to be true for
the RSJ dynamics with PBC ($z\approx 1.6$ at $T=0.90$). Our
conclusion is that the dynamical critical exponent $z$ is a
boundary sensitive quantity. 
We also note that the FTBC is adequate for describing an open system with voltage
fluctuation across the system and that consequently the $z$ values
obtained for this case describe the most usual physical situation.

\begin{figure}
\centerline{\epsfxsize=8.0cm \epsfbox{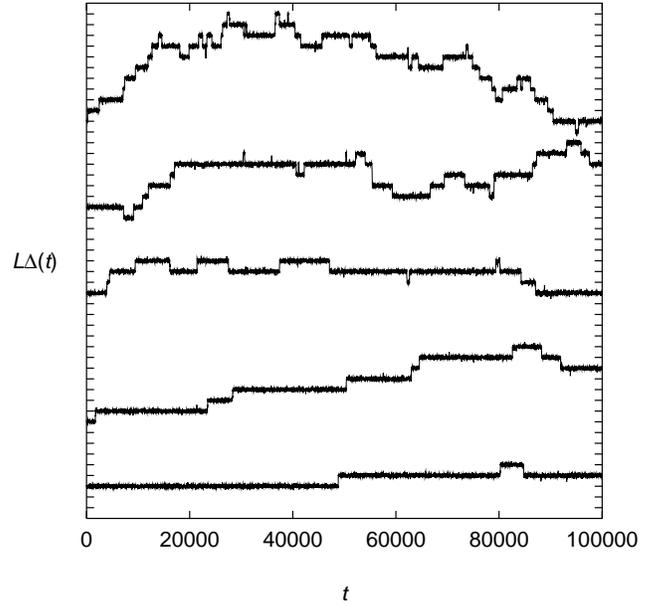}}
\vskip  -2cm
\caption{Time evolution of the variable $L\Delta(t)$ at $T=0.85$ as a function of time $t$ 
for RSJ dynamics and system sizes $L$ = 6, 8, 10, 12, and 16 (from top
to bottom). The curves are shifted in the vertical direction. 
As seen $L\Delta(t)$ sometimes makes discrete jumps
of size $2\pi$ (the unit of the vertical axis is $2\pi$). 
The characteristic time $\tau$ in Fig.~\ref{fig9} is related to the average 
time between the $2\pi$ jumps.
}
\label{fig8}
\end{figure}

\begin{figure}
\centerline{\epsfxsize=9.0cm \epsfbox{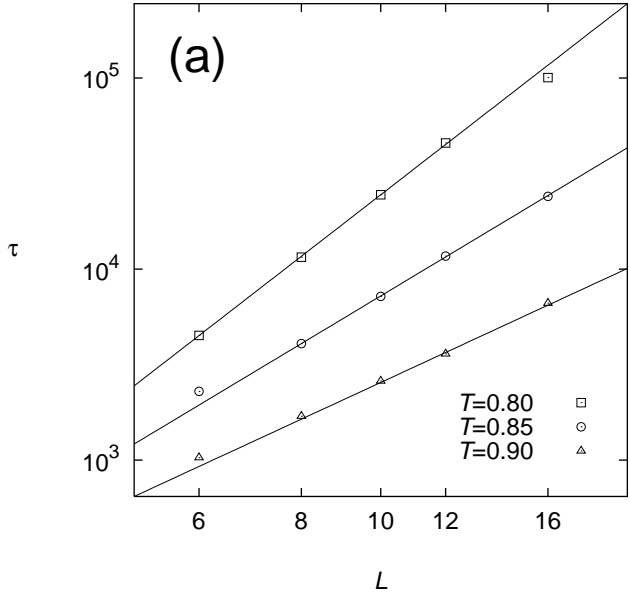}}
\vskip -4cm
\centerline{\epsfxsize=9.0cm \epsfbox{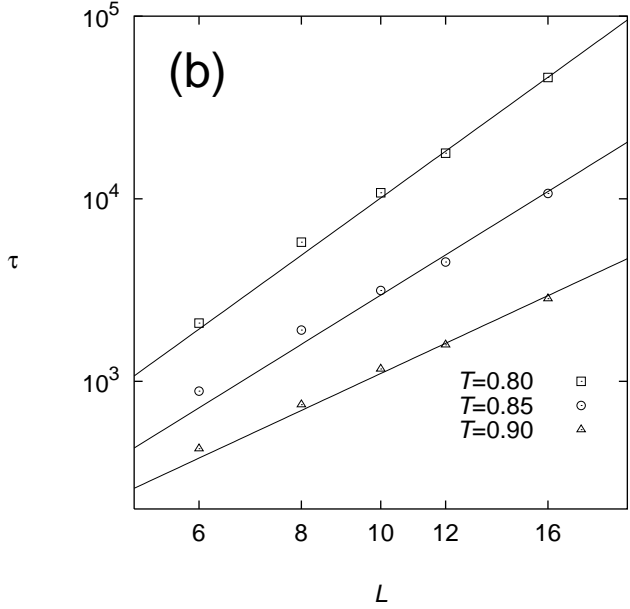}}
\vskip  -2cm
\caption{The relaxation time $\tau$ obtained directly from the time
  scale of the $2\pi$ jumps of
  $L\Delta(t)$. The obtained values of $\tau$ are  
plotted against the system size $L$ for (a) RSJ  and (b) TDGL dynamics (see Fig.~\ref{fig8}).
The full lines represent $\tau \sim L^{z}$ with 
the $z$ values taken from Fig.~\ref{fig7}.
The figure illustrates that $z$ determined from the scaling of the
resistance $R$ is indeed associated with a divergent characteristic time.
}
\label{fig9}
\end{figure}

It is in fact possible to estimate the characteristic time $\tau$ very
directly since the variable $\Delta_x$
changes by the amount $2\pi/L$ when a vortex moves across the system
in the $y$ direction, as discussed in Sec.~\ref{sec_dyn}. Every such event consequently is signaled by a
$2\pi$ step in the time series of the variable $L\Delta_x$. Figure~\ref{fig8}
illustrates this for the RSJ dynamics at $T=0.85$ for various system
sizes. As seen in the figure the $2\pi$ jumps are very well observable. The
characteristic time scale $\tau$  of these 
$2\pi$ jumps is easily estimated as the average time
between the jumps and we expect that $\tau \sim L^z$ with the same dynamical
critical exponent $z$ as in $R \sim L^{-z}$. Figure~\ref{fig9}(a) shows $\tau$ 
plotted against the system size $L$ in a log-log plot for the
RSJ dynamics for three different temperatures (in practice we
use a coarse graining of 100 time units in our estimate of the average time between the 
$2\pi$ jumps). The full drawn
straight lines in Fig.~\ref{fig9}(a) have the slopes given by the $z$ values determined 
previously from the calculation of $R$ [see  Fig.~\ref{fig7}(a)]. As seen the two
ways of determining $z$ agree very well. Figures~\ref{fig7}(b)
and \ref{fig9}(b) illustrate the same agreement in case of
TDGL dynamics.

\begin{figure}
\centerline{\epsfxsize=8.0cm \epsfbox{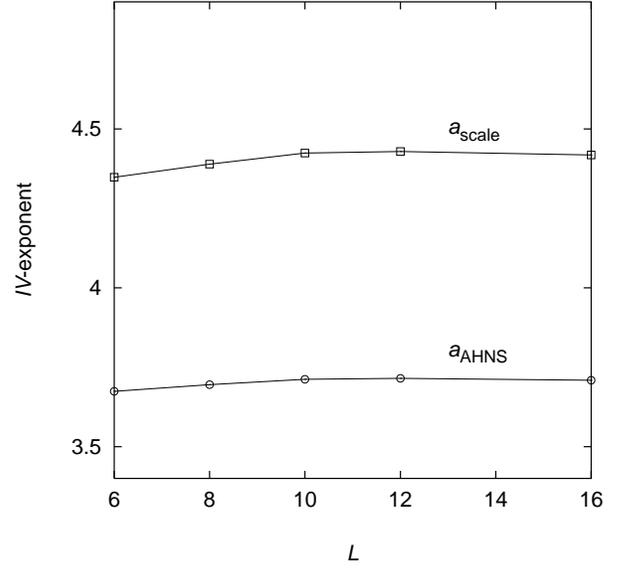}}
\vskip  -2cm
\caption{Predictions of the $IV$ exponent for the RSJ model
at $T=0.8$ as a function of the system size $L$. 
The open squares are obtained from
$a_{\rm scale} =  1/\tilde\epsilon T^{\rm CG}_{\rm eff} - 1$
for FTBC whereas the open circles represent
$a_{\rm AHNS} = 1/2\tilde\epsilon T^{\rm CG}_{\rm eff} + 1$
for FTBC.
}
\label{fig10}
\end{figure}

Let us now  consider what happens when a finite current is applied
across the system. The scaling argument by Dorsey~\cite{dorsey} makes
use of the fact that the current density ${\mathcal{J}}$ introduces
a new length scale $1/{\mathcal{J}}$.~\cite{fisher} This new length scale
replaces the finite size $L$ in the leading $L$ dependence of $R$, so
that~\cite{foote}
\[
V=R({\mathcal{J}}^{-1})I\propto \frac{1}{\tau({\mathcal{J}}^{-1})}I\propto
  {\mathcal{J}}^zI
  \]
and consequently
 $ V\propto I^{z+1}$
below the KT transition as suggested in Ref.~\onlinecite{minnhagen1}.
From the finite-size scaling of $R$ and $\tau$ we obtain $z$ and using the 
scaling argument this $z$ is related to the nonlinear $IV$ exponent by $a=z+1$ where
$V \sim I^a$. In Table~\ref{tabI} we have given the values of  $a=z+1$,
 where $z$ has been obtained from the finite-size scaling of $R$.  
Another scaling argument~\cite{minnhagen1} gives [see Eq.~(\ref{eq_zscaling})]
$z = 1/\tilde\epsilon T^{\rm CG} - 2$ and consequently $a=z+1=1/\tilde\epsilon T^{\rm CG} - 1$.
In order to compare this scaling prediction with the $z$ values obtained directly from 
the finite-size scaling of $R$, we need to estimate
$1/\tilde{\epsilon}T^{\rm CG}$. As described in Sec.~\ref{sec_XY}, $T^{\rm CG}$
  is given by $T^{\rm CG}=T/(2\pi J\langle U''\rangle)$ and
  $1/\tilde{\epsilon}=\lim_{k\rightarrow 0} 1/\hat{\epsilon}(k,0)$.
  However for FTBC we have
\[ \frac{1}{\tilde{\epsilon}}=\lim_{k\rightarrow 0} \frac{1}{\hat{\epsilon}(k,0)} >
\frac{1}{\hat{\epsilon}(0,0) }=0 . \]
So for each size $L$ we estimate $1/\tilde{\epsilon}$ by
$1/\hat{\epsilon}(2\pi/L,0)$.
As mentioned in the beginning of this section we can also include a small correction due to the finite
time step $\Delta t$ in the simulations for each size $L$ by
replacing $T$ by an effective temperature
$T^{\rm eff}=\hat{G}(0,0)/\langle U'' \rangle$. Figure~\ref{fig10} shows 
$a_{\rm scale} = z + 1 = 1/\tilde\epsilon T^{\rm CG} - 1$ estimated in this way as a
function of $L$. When comparing with the $a$ values obtained from the
finite-size scaling of $R$, we take an average over the relevant
$L$ interval. These values are shown in Table~\ref{tabI}. As is
apparent from Table~\ref{tabI}, the values of $a$ determined from the size
scaling of $R$ and $\tau$ agree very well with $a_{\rm scale}$ both for the
RSJ case and the TDGL case. 
Thus we conclude that $z=(1/\tilde{\epsilon}T^{\rm CG})-2$.
This conclusion has also been reached for
the lattice Coulomb gas model with Monte Carlo dynamics.~\cite{weber}
Furthermore, by invoking the scaling argument described above, we
infer that
the $IV$ exponent should be given by
$a=a_{\rm scale}=z+1=1/\tilde{\epsilon}T^{\rm CG}-1$.~\cite{minnhagen1}

The model given in Ref.~\onlinecite{bormann} suggests the
finite-size scaling $R\propto L^{1-a_{\rm scale}}$ in agreement with
our results.~\cite{bormannprivate} 
However, according to the reasoning in
Ref.~\onlinecite{bormann}, the scaling argument $L\propto 1/{\mathcal{J}}$
leading to the nonlinear $IV$ exponent $a=a_{\rm scale}$
should break down for small enough currents and in this limit one
should instead recover $a=a_{\rm AHNS}$.

In the next section we investigate the nonlinear $IV$ characteristics
more directly by imposing an external current.

\section{Nonlinear $IV$ Characteristics} \label{sec_nonlinear}

\begin{figure}
\vskip  -2.5cm
\centerline{\epsfxsize=8.0cm \epsfbox{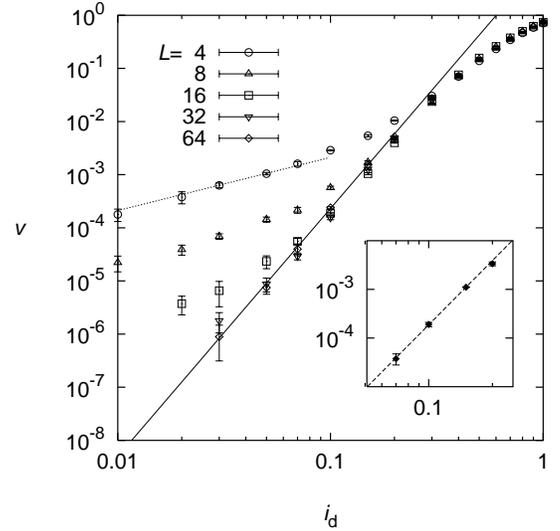}}
\vskip  -2.0cm
\caption{The current-voltage ($IV$) characteristics at $T=0.8$ for the fluctuating
twist boundary condition at various system sizes. 
The full straight line is 
obtained from the least-square fit in the interval $0.03 \le i_d \le 0.15$ for $L=64$
 which gives $a \approx 4.7$. 
The linear region with $IV$ exponent
$a = 1$, seen for the smaller sizes and small currents (the dotted straight line
has the slope $a=1$), disappears as
the system size is increased.
Inset: $IV$ curve for
$L=64$ at fixed Coulomb gas temperature $T^{\rm CG} \approx 0.17$, corresponding
to $T=0.80$ with no external currents. The broken line is obtained
from the least-square fit in the interval $0.07 \le i_d \le
0.15$, giving  $a \approx 4.5$. }
\label{fig11}
\end{figure}

In order to obtain the $IV$ characteristics for the 2D $XY$ model
with RSJ dynamics we use FTBC and Eqs.~(\ref{eq_Deltax_Id}) -- (\ref{eq_Vx}). Figure~\ref{fig11} 
shows the data obtained from lattice sizes $L=4$ to 64,  where
$v=V/L$ is plotted against $i_d=I_d/L$ in a log-log plot.
As seen from the figure the data are size dependent but for $L=64$
the data appear to be reasonably size converged except
for the smallest currents. The data in the figure are for $T=0.80$ and
the straight line is a least-square fit to the $L=64$ data in the current interval
$0.03\leq i_d\leq 0.15$ and gives
$a\approx 4.7$, which is in reasonable agreement with 
$a_{\rm scale} = 1/{\tilde \epsilon}T^{\rm CG} - 1 \approx 4.5$. 
In the following we will investigate the sensitivity of
 this apparent agreement to finite size, finite current, and boundary conditions.

One finite current effect is that the exponent $a$
refers to a constant Coulomb gas temperature
$T^{\rm CG} = T/[2\pi J \langle U'' \rangle ]$. Since a 
finite current changes the value of
$\langle U'' \rangle$,~\cite{minnhagen1}
fixed temperature ($T$ = const) is not entirely
equivalent to fixed Coulomb gas temperature ($T^{\rm CG} = $ const).
In order to convert the data to fixed Coulomb gas temperature 
we have calculated $v$ and $T^{\rm CG}$ for $T=0.79$ and $0.80$
for fixed 
external currents, and then by interpolation estimated 
the voltage value corresponding to
a fixed $T^{\rm CG}$. The resulting data for a fixed
Coulomb gas temperature ($T^{{\rm CG}}\approx 0.17$)
are shown in the inset of Fig.~\ref{fig11}.
The broken line in the inset is a fit to the data and
gives $a\approx 4.5$. Thus this correction leads
to a somewhat smaller value of $a$.

\begin{figure}
\centerline{\epsfxsize=8.0cm \epsfbox{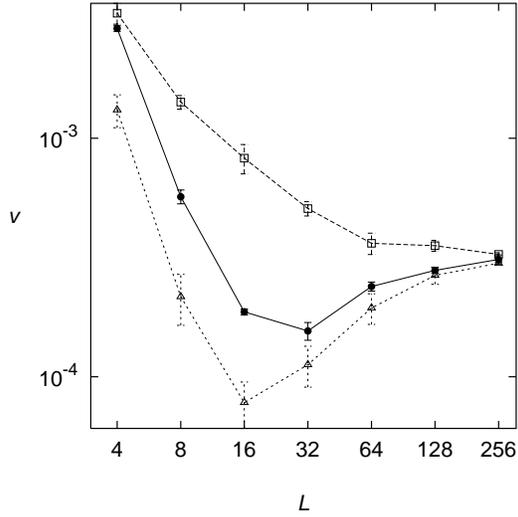}}
\vskip  -2cm
\caption{Voltage $v$ versus system size $L$ at the current 
$i_d = 0.1$ for $T=0.8$. The empty squares are for the uniform current
injection with periodic boundary conditions in the direction perpendicular to
the current. The empty triangles are obtained with the critical current $i_c = 10$
for vertical junctions on the boundaries, which is very similar to the busbar boundary.
The filled circles are for FTBC introduced in Sec.~\protect\ref{sec_dyn}. As the system size is increased, 
the voltages for all
three methods are shown to converge towards the same value in the $L=\infty$ limit.
However, the uniform current injection approaches the $L=\infty$ limit
from above whereas the FTBC and busbar condition approach from below. 
The lines are guides to the eye.}
\label{fig12}
\end{figure}

All previous estimates for the nonlinear $IV$ exponent for
the RSJ model have been obtained for $L=64$ or smaller
sizes.~\cite{simkin,minnhagen1,mon} The next question we address is how 
much the possible remaining size effects could alter the results inferred for $L=64$.
Figure~\ref{fig12} shows voltage $v$ versus the system size $L$ 
at the external current $i_d = 0.1$ and $T=0.8$
for three different cases. 
The open squares at the top correspond 
to the usual uniform current injection method used in
Ref.~\onlinecite{mon}.
The filled circles correspond to our FTBC boundary
condition and finally the open triangles 
at the bottom correspond
to the busbar boundary condition used in Ref.~\onlinecite{simkin}.~\cite{comsimkin}
It is clear from the figure that $L=\infty$ result
cannot be estimated by the $L=64$ for $i_d=0.1$.
For smaller $i_d$ the situation quickly gets even worse.
Thus this unexpected strong size dependency clearly makes all
earlier results obtained for $a$ from $IV$ simulations
somewhat questionable.~\cite{simkin,minnhagen1,mon}

As seen in Fig.~\ref{fig12} the uniform current injection appears
to approach the $L=\infty$ value from above whereas the FTBC and the busbar
condition appear to approach the $L=\infty$ value from below.
We have found this to be generally true. 
From this we conclude that $L=256$ is enough to estimate the $L=\infty$ limit for $i_d>0.1$, since 
the data for FTBC and uniform current injection are closely
the same in this case. The value of $a$ obtained in this converged 
current region is about $a\approx 4.1$, which
is somewhat smaller than $a\approx 4.3$ obtained from the finite-size scaling
of $R$ in the previous section.

\begin{figure}
\centerline{\epsfxsize=8.0cm \epsfbox{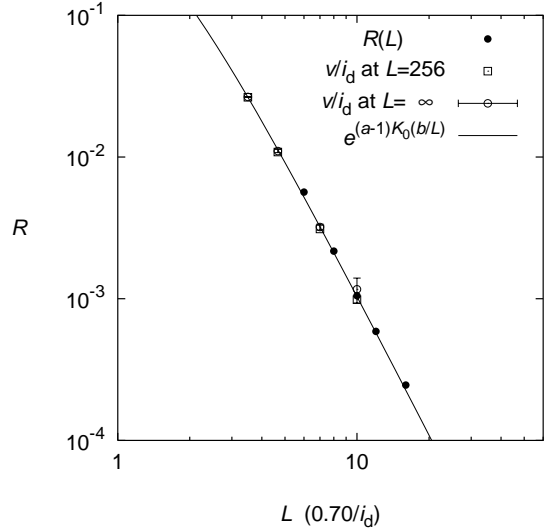}}
\vskip  -2cm
\caption{The resistance $R(c/i_d)=v/i_d$ at $T=0.80$ (open squares correspond to
the $L=256$ data for FTBC and open circles to the average between the
$L=256$ data for FTBC and the uniform current injection) is compared to 
the resistances $R(L)$ at $T=0.80$ [filled circles, the same data as in
Fig.~\ref{fig7}(a)].
Choosing the constant $c\approx 0.70$ makes the two data sets collapse
onto a single curve. The full drawn curve interpolates between
the limits $R=1$ for large currents and $R\propto (i_d)^{a-1}$ for small
currents (the explicit form of the interpolation curve is
$e^{(a-1)K_0(b/L)}$ with $a=4.3$ and $b=1.42$). The figure suggests
that the two ways of calculating $R$ are consistent and
that the $R(c/i_d)$ data are not quite in the asymptotic
small-current regime. 
}
\label{fig13}
\end{figure}

In order to get some further insight, we note that the present
 simulation gives the resistance
 $R=v/i_d$ as a function of $i_d$,
as discussed in the previous section, for small enough current densities ${\mathcal{J}}$, 
$1/{\mathcal{J}}$ should corresponds to a finite $L$. Consequently
$R(c/i_d)$, where $c$ is a constant, obtained in the present simulations should
 be equivalent
 to $R(L)$ obtained in the previous section:
For an appropriate choice of the constant $c$ the data for these two simulations should
fall on a single curve.
Figure~\ref{fig13} illustrates this equivalence,
the filled circles are the data for $R(L)$ and the open squares are the $IV$ data obtained from FTBC with $L=256$.
The open circles are the averages between the $L=256$ result for FTBC
and uniform current injection. When the open circles and squares overlap,
the $L=\infty$ limit has been reached.
As seen from the figure the two data sets for $R$ to a good
approximation fall on a single curve. For large currents $R$
approaches the junction resistance $r=1$ and for small currents
$R\propto (i_d)^{a-1}$.
The full drawn curve ($R=e^{(a-1)K_0(bi_d)}$ where $K_0$ is a modified Bessel
 function) interpolates between these two limits [$K_0(x) \sim - \ln x$ for small
$x$ and $K_0(0) = 0$].
 Since the converged $IV$ data are higher up on the curve one
 expects an apparent smaller $a$ than for the $R(L)$ data which
 are lower down on the curve. Our conclusion is that the results
 from the $IV$ simulations and the $R(L)$ simulations are 
consistent with each other and with the scaling assumption.  

\subsection*{Scaling collapse}

\begin{figure}
\centerline{\epsfxsize=8.0cm \epsfbox{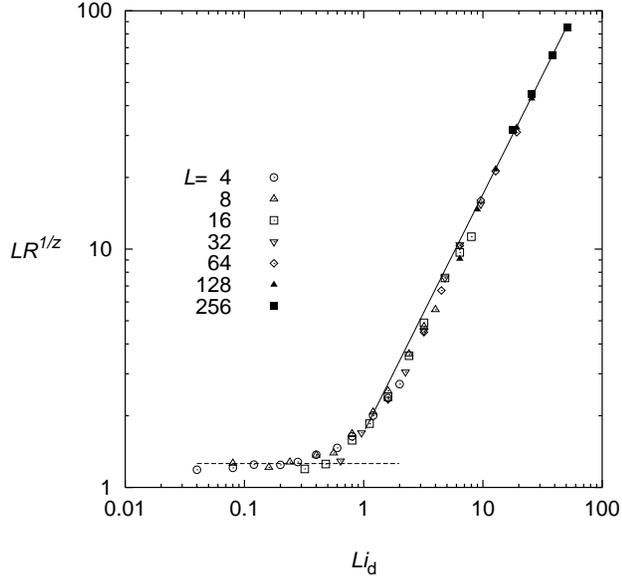}}
\vskip  -2cm
\caption{Demonstration of the validity of the scaling assumption. The
  $IV$ data for the RSJ model with $T=0.8$ and $i_d\leq 0.6$
  are plotted as $LR^{1/z}$ against $Li_d$. For $z\approx
  3.3$ all the data for the various $L$ and 
  $i_d$ collapse onto a single scaling function $f(x=Li_d)$. 
The horizontal broken line corresponds to the constant value
  for $LR(L)^{1/z}$ obtained for $i_d=0$ for
  the same value of $z$ [see  Fig.~\ref{fig7}(a)]. The straight line 
corresponds to the linear behavior $f(x) \sim x$ for large $x$.}
\label{fig14}
\end{figure}

It is in fact possible to demonstrate the validity of the scaling
assumption in a more general way: At fixed temperature $R$ is only a function of $L$
and ${\mathcal{J}}$. From the fact that $R \sim 1/L^z$ at ${\mathcal{J}} = 0$ and
that the combination ${\mathcal{J}}L$ is
dimensionless,  one expects that
\begin{equation}
R=\left[ \frac{f({\mathcal{J}}L)}{L}\right]^z  ,
\label{scaleR}
\end{equation}
where $f(x)$ is a dimensionless scaling function. The scaling function
$f(x)$ must have the limits $f(0)=$const since $R \sim 1/L^z$ for ${\mathcal{J}} = 0$,  
and $f(x)\propto x$ 
for large $x$. The latter follows because the $L\rightarrow \infty$ limit
has to give a nonvanishing finite $R$. This means that the combination 
\begin{equation}
LR^{1/z}=f({\mathcal{J}}L)
\label{scalecomb}
\end{equation}
is only a function of ${\mathcal{J}}L$. In Fig.~\ref{fig14} we have
plotted all our simulation data for $i_d\leq 0.6$ as
$LR^{1/z}$ against $i_dL$, i.e., the data
shown in Fig.~\ref{fig11}
together with  data for $L=128$ and 256. 
Using $z$ as an adjustable parameter, we find that all the
data collapse onto a single scaling curve for $z\approx 3.3$. We
emphasize that this scaling collapse involves only {\em one}
free parameter, $z$. One also notes that the best value for the
collapse (obtained by a least-square method) is closely the same
($z\approx 3.3$ at $T=0.80$) as was found in the absence of external currents
shown in Fig.~\ref{fig7}(a). 
Furthermore, this zero-$i_d$ data collapse onto a
single value for $z\approx 3.3$ when plotted as $LR^{1/z}$ and
this constant value is given by the broken horizontal line in
Fig.~\ref{fig14}. Thus the data collapse shown in Fig.~\ref{fig14}
clearly demonstrates that the scaling assumption is valid for all the
data we have obtained. Since the scaling assumption gives
$a=a_{\rm scale}=z+1=1/\tilde{\epsilon}T^{\rm CG}-1$, our conclusion is
that $a_{\rm scale}$ is indeed the correct $IV$ exponent over a broad
parameter range.

The model discussed in Ref.~\onlinecite{bormann} suggests that for small
enough $i_d$ the scaling assumption should break down. Thus for
such small currents the data for large enough $L$ should fall above
the scaling curve in Fig.~\ref{fig14}. There is no sign of any such
deviation in our data. However, this does not preclude the possibility
that such a deviation could in principle occur for larger sizes and smaller
currents than we have been able to investigate.
 
\begin{figure}
\vskip  -2cm
\centerline{\epsfxsize=8.5cm \epsfbox{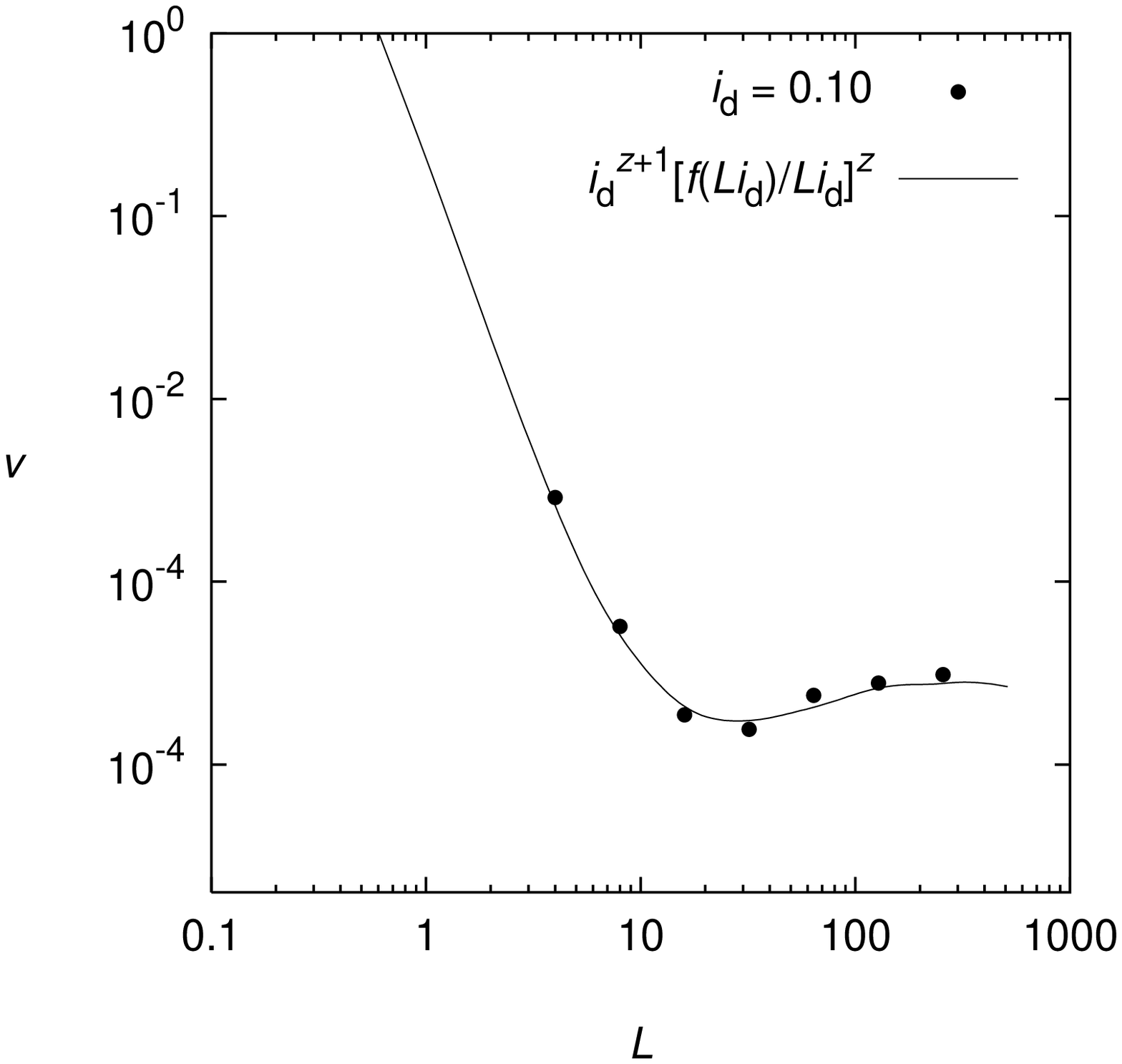}}
\vskip  -2cm
\caption{The relation between the finite-size dependence of the voltage $v$ and
  the scaling function $f(x=Li_d)$. The full drawn curve is the
  function $v=i_d^{z+1}[ f(x)/x ]^z$ where
  $f(x)$ has been obtained by a data smoothing of the data in
  Fig.~\ref{fig14}.
The filled circles are the finite-size data for $v$
  at $T=0.8$, the same data as the filled circles in Fig.~\ref{fig12}.}
\label{fig15}
\end{figure}

It is also interesting to note that the scaling function $f(x)$ is
intimately connected to the finite-size dependence of the voltage for FTBC.
[See, for example,  Fig.~\ref{fig12} for $T=0.8$ and $i_d=0.1$
(filled circles).] According to Eq.~(\ref{scaleR}) we have
\begin{equation}
v=i_d^{z+1}\left[\frac{f(Li_d)}{Li_d}\right]^z .
\label{iscale}
\end{equation}
The full drawn curve in Fig.~\ref{fig15} gives $v$ as a function of $L$
using Eq.~(\ref{iscale})
for $i_d=0.1$ where the scaling function $f(x)$ has been
obtained by a data smoothing of the data in Fig.~\ref{fig14}.
The filled circles
is a replot of the finite-size dependence given as filled circles in Fig.~\ref{fig12}. As
is apparent from Fig.~\ref{fig15}, the
particular shape of the finite-size dependence is a direct reflection of the
scaling function $f(x)$.

The AHNS prediction~\cite{ambegaokar} for the nonlinear $IV$ exponent differs from
the scaling prediction and is instead given by
\[
a_{\rm AHNS} = \frac{1}{2\tilde\epsilon T^{\rm CG}} +1 .
\]
The corresponding values are given in Table~\ref{tabI} and Fig.~\ref{fig10}. Our
simulations support the scaling prediction. E.g., for $T=0.8$ and RSJ we find
$a \approx 4.3$ which is close to the scaling prediction $a_{\rm scal} \approx 4.4$ and
differs from the AHNS prediction $a_{\rm AHNS} \approx 3.7$. 

\section{Summary and Comparisons} \label{sec_final}

The first main result of the present investigation is that the 2D $XY$ model
with RSJ dynamics is well described by the MP form for the dynamical
response.
This appears to be generic for 2D vortex
fluctuations since the same form has been found for the $XY$ model with
TDGL dynamics,~\cite{jonsson} the 2D Coulomb gas with Langevin dynamics~\cite{holmlund} as
well as in experiments.~\cite{jonsson,theron,rogers} However, 
since the 2D $XY$ model with RSJ dynamics is generally 
accepted as a valid model for a 2D Josephson junction array, the present investigation ties the MP form found in the present and
previous simulations
closer to the MP form found 
in experiments.~\cite{jonsson,theron}

We found the critical exponent $z=2$ at the KT transition from the finite-size
scaling of the resistance $R$ using the fluctuating twist boundary condition
FTBC, both in case of RSJ and TDGL dynamics. 
Furthermore, we found the same value of $z$ for RSJ and TDGL for all temperatures
below the transition using the same method.
However, we also found that the finite-size scaling with PBC gave different results.
Thus it appears as if the finite-size scaling determination of $z$ depends on the boundary
condition. Our conclusion is that it fails for PBC because the characteristic time $\tau$
is inversely proportional to the resistance $R$ and for PBC the resistance $R$ is identically
zero for any finite size. This suggests that the proper value of $z$ cannot be determined from
finite-size scaling with PBC.

The exponent $z$ determined from the finite-size scaling with FTBC were found to be the
same for RSJ  and
TDGL dynamics and to support the scaling prediction
$z=1/\tilde{\epsilon}T^{\rm CG}-2$ in agreement with what was found in
Ref.~\onlinecite{weber} for the 2D lattice Coulomb gas with Monte Carlo
dynamics. We also explicitly showed that the exponent $z$ determined
from the finite-size scaling of $R$ is related directly to a diverging relaxation
time. Thus our conclusion is that $z$ is larger than  2 below the
KT transition. This result is in agreement with the model discussed in
Ref.~\onlinecite{bormann}.~\cite{bormannprivate}  
Using a scaling argument,~\cite{dorsey} we related the finite-size scaling of
$R$ to the nonlinear $IV$ characteristics by noting that the current
density ${\mathcal{J}}$ plays the role of $1/L$ leading to $V\propto
I^a$ with $a=z+1$. Consequently, provided the scaling argument is
valid, our simulations support the
prediction $a=1/\tilde{\epsilon}T^{\rm CG}-1$.~\cite{minnhagen1}

We also calculated the $IV$ exponent $a$ directly from 
the voltage $V$ as a function of current $I$. 
Here we found that the results were strongly size dependent. This large size dependence 
we found for standard current injection boundary, FTBC, and
the ``busbar'' boundary condition introduced in Ref.~\onlinecite{simkin}. For our largest lattice
sizes $256\times 256$ a size-converged result
could only be estimated for currents which seemed to be 
outside the true scaling regime $V\propto I^a$. However, by using
the relation $L\propto 1/{\mathcal{J}}$ valid for small enough
${\mathcal{J}}$ we showed that the data for the resistance simulation
  $R(L)$ and the $IV$ simulations $R(c/{\mathcal{J}})$ can be made to
fall on a
    single curve for an appropriate choice of the constant $c$. This
    agreement suggests that our $IV$ simulations and our
    $R(L)$ simulations are consistent with each other and with the scaling assumption. We concluded
 that it is difficult to obtain the nonlinear
 $IV$ exponent $a$ directly from the $V(I)$ data in case of the 2D $XY$ model with RSJ dynamics.
 This is because resistance ratios $R(I)/r<10^{-3}$ ($r$ is the
 junction resistance) seem to be needed. This in turn implies such small currents that
 lattice sizes considerably larger than $256\times 256$ are required
 to avoid the finite-size effects.
However, in case of the 2D Coulomb gas with
    Langevin dynamics~\cite{holmlund} it has been possible to converge
    the simulations closer to where the true scaling $V\propto I^a$ appears to be 
    valid and in these cases the scaling exponent
    $a=1/\tilde{\epsilon}T^{\rm CG}-1$ was deduced from the $V(I)$ data.

Finally, we showed that all our $IV$ data and our $R(L)$ data for a fixed temperature
collapse onto a single scaling curve $f(x=Li_d)$. This data
collapse demonstrates that the scaling argument is indeed valid over
a broad parameter range and thus confirms that the nonlinear
$IV$ exponent is given by $a_{\rm
scale}=1/\tilde{\epsilon}T^{\rm CG}-1$ over the parameter range
covered by our data. This does not
preclude the possibility that, for smaller currents and larger sizes
than we have been able to converge, there might be a deviation from the
scaling curve given in Fig.~\ref{fig14} as suggested by the model in Ref.~\onlinecite{bormann}.
However, there is no sign of
any deviation from the scaling curve in our data for the
RSJ model.

In short, the present simulations of the 2D $XY$ model with RSJ dynamics
confirm the picture that 2D vortex fluctuations has an anomalous kind of
dynamics. The characteristic features of this dynamics are presumably
linked to the logarithmic vortex interaction.
However, a firmer theoretical understanding of the characteristic
features, which have been encountered in numerous simulations, as well as in
experiments, is still lacking and is a challenge for future research.

\acknowledgments
One of the authors (B.J.K.) wishes to acknowledge
the financial support of Korea Research Foundation
for the program year 1997. The research was supported by the Swedish
Natural Research Council through Contract Nos. FU 04040-332 and EG 10376-310. 

\end{multicols}
\noindent\rule{0.5\textwidth}{0.1ex}\rule{0.1ex}{2ex}\hfill
\widetext
\appendix
\section*{Linear response} \label{app_a}

A total current
$i_x({\bf r}, t)$ which varies slowly in time compared to the thermal
fluctuations gives rise to an average nonvanishing phase difference
$q({\bf r}, t)=\langle \nabla_x \theta ({\bf r}, t)\rangle$. Thus  Eqs.~(\ref{ix})
and (\ref{P}) together with the chain rule gives
\begin{equation}
  \dot{P}({\bf r}-{\bf r}', t-t')=-J\int d^2r''dt''\left. 
\frac{\partial \langle U'[\nabla_x\theta ({\bf
      r}, t)]\rangle }{\partial q({\bf r}'',t'')}\right|_0
  \cdot\left. \frac{\partial \langle \nabla_x\theta({\bf
      r}'',t'')\rangle}{\partial i_x({\bf r}', t')}\right|_0
  + \delta({\bf r}-{\bf r}')\delta(t-t')  ,
  \label{P2}
\end{equation}
where $|_0$ denote that the resulting averages should be the equilibrium ones.
Let us introduce the notation
\[
Q({\bf r}-{\bf r}'', t-t'')= J\left.\frac{\partial \langle U'[\nabla_x\theta ({\bf
      r}, t) ] \rangle }{\partial q({\bf r}'',t'')}\right|_0
\]
then the Fourier transform of Eq.~(\ref{P2}) is just
\begin{equation}
i\omega \hat{P}({\bf k},\omega )=
-\hat{Q}({\bf k},\omega )\hat{P}({\bf k},\omega )+1
\label{PQ}
\end{equation}
so that 
\[
\hat{P}({\bf k},\omega )=\frac{1}{i\omega +\hat{Q}({\bf k},\omega )}  .
\]
We note that 
\begin{eqnarray}
Q({\bf r}-{\bf r}', t-t') &=& J\left. \frac{\partial \langle U'[\nabla_x\theta ({\bf
      r}, t) ] \rangle }{\partial q({\bf r}',t')}\right|_0 \nonumber \\
&=&J\langle U''[\nabla\theta({\bf r},t) ] \rangle
  \delta ({\bf r}-{\bf r'})\delta (t-t') + J\left.\frac{\partial \langle U'[\nabla_x\theta ({\bf
      r}, t) ] \rangle }{\partial q({\bf r}',t')}\right|_0  . \nonumber
\end{eqnarray}
Here the last term is for $t\neq t'$ and $r\neq r'$ so that the
 disturbance
 $q({\bf r}', t')=\langle \nabla_x\theta ({\bf r}',t')\rangle$ couples linearly to
 $JU'[\nabla_x\theta ({\bf r}',t')]$ in the $XY$ Hamiltonian. Consequently, the corresponding correlation function is
\[-J^2\langle U'[\nabla_x\theta ({\bf r},t)] 
U'[\nabla_x\theta ({\bf r}',t')]\rangle\]
and by the fluctuation-dissipation theorem we have
\[
Q({\bf r},t)=\frac{J^2}{T}\frac{\partial}{\partial t}\langle U'[\nabla_x\theta ({\bf r},t)] 
U'[\nabla_x\theta (0,0)]\rangle +J\langle U''[\nabla_x\theta(0,0)]\rangle  \delta ({\bf r})\delta (t)
\]
for $t\geq 0$ and $0$ otherwise. Next we note that a space Fourier transform of the correlation
function $J^2\langle$ $U'[\nabla_x\theta ({\bf r},t)]$ $U'[\nabla_x\theta ({\bf r}',t')]\rangle$ gives the correlation function
$\hat{G}({\bf k}, t)$ defined in connection with Eq.~(\ref{G}) so that
\begin{eqnarray}
  \hat{Q}({\bf k}, \omega ) &=&
  \int_0^\infty dt e^{-i\omega t}\hat{Q}({\bf
  k},t)=\rho_0+\frac{1}{T}\int_0^\infty dt e^{-i\omega
  t}\frac{\partial}{\partial t}\hat{G}({\bf k}, t) \nonumber \\
  & = & \rho_0 -\frac{1}{T}\hat{G}({\bf k},0)-\frac{1}{T}\int_0^\infty dt
  e^{-i\omega t}\hat{G}({\bf k}, t)=
  \frac{\rho_0}{\hat{\epsilon}({\bf k},\omega )} ,
\end{eqnarray}
where $\rho_0=J\langle U''\rangle$ and the result is obtained by
partial integration and comparison with Eqs.~(\ref{reps})--(\ref{eps}).

\begin{multicols}{2}
\narrowtext

\begin{table}
\caption{ Comparison between the exponent $a\equiv z+1$ obtained from
  the $R(L)$ simulations and the  predicted values
$a_{\rm scale}$ and $a_{\rm AHNS}$ for RSJ and TDGL dynamics.
The values of $a_{\rm scale}$ and $a_{\rm AHNS}$ are obtained from the averages
over $L=$ 10, 12, and 16 (see Fig.~\ref{fig10} for RSJ case). 
The exponent $a\equiv z+1$ is obtained from $R(L) \sim L^{-z}$ in Fig.~\ref{fig7}
and is found to be consistent with $z$ in $\tau \sim L^z$ in Fig.~\ref{fig9}.
The numbers in parentheses represent the statistical
errors of the last digits. It is clearly shown that the exponent $a$
measured by direct calculation of resistance from Eq.~(\ref{eq_R}) is much 
closer to $a_{\rm scale}$ than to $a_{\rm AHNS}$ 
for {\em both} RSJ and TDGL dynamics.}
\vskip 1cm

\label{tabI}
\begin{tabular}{r r c l l}
$T$  & $a_{\rm scale}$ &      &  $a_{\rm AHNS}$ & a \\ \hline 
     &                 & RSJ  &                 &  \\
0.80 &        4.42(2)  &      &  3.71(2)        & 4.3(1) \\
0.85 &        3.80(2)  &      &  3.40(2)        & 3.7(1) \\
0.90 &        3.05(2)  &      &  3.02(2)        & 3.0(1)   \\
     &                 & TDGL &                 &  \\
0.80 &        4.55(3)  &      &  3.77(2)        & 4.3(1) \\ 
0.85 &        3.85(2)  &      &  3.43(2)        & 3.8(1) \\ 
0.90 &        3.12(2)  &      &  3.06(1)        & 3.1(1)  
\end{tabular}
\end{table}

\newpage

\newpage

\newpage

\end{multicols}
  

\begin{thebibliography}{99}
\bibitem{foota} We define the term boundary condition as any imposed constraint whose effect
vanishes for $L = \infty$. Note that this definition includes constraints  on the dynamics
for a finite system provided this constraint has no effect for $L=\infty$.
\bibitem{footc} In practice we use integration times corresponding to about $10^7$ time steps.
In practice this gives us a smallest converged frequency of about $\omega \approx 0.05$.
\bibitem{footd} PBC restricts the resistance $R$ to be identically zero for all lattice
sizes. We believe that the characteristic time $\tau$ which is relevant to the critical
exponent $z$ is inversely proportional to $R$. This means that the relevant characteristic
time is infinite for all finite lattice sizes in case of PBC so that the correct $z$ cannot
be inferred from finite lattice sizes in case of PBC.
\bibitem{comsimkin} For convenience we used an approximate version
  of the busbar idea by using uniform current injection but with the
  coupling between the junctions at the boundary ten times larger than
  the others.
  We found that this procedure closely reproduced the strict busbar condition.
\end{thebibliography}
\end{document}